\documentclass[12pt]{iopart}
\usepackage{amstext}
\usepackage{iopams}  
\usepackage[sort&compress,comma]{natbib}

\usepackage[utf8]{inputenc}
\usepackage{booktabs}
\usepackage{ulem}
\usepackage{xcolor}
\usepackage{amsmath,bm, amssymb}
\usepackage{graphicx}
\usepackage[labelfont=bf]{caption}
\usepackage{float}
\usepackage[yyyymmdd]{datetime} 
\usepackage{siunitx}
\usepackage{hyperref}

\begin{document}
\title[Convolutive BSS for identifying spiking motor neuron activity]{Revisiting convolutive blind source separation for identifying spiking motor neuron activity: From theory to practice}

\author{Thomas Klotz$^{1,*}$ and Robin Rohl{\'e}n$^{2,*}$}

\address{$^1$ Institute for Modelling and Simulation of Biomechanical Systems, University of Stuttgart, Pfaffenwaldring 5a, 70569 Stuttgart, Germany\\[2mm]
$^2$ Department of Diagnostics and Intervention,
Ume{\aa} University, 901 87 Ume{\aa}, Sweden}
\ead{thomas.klotz@imsb.uni-stuttgart.de and robin.rohlen@umu.se}

\vspace{8pt}
\begin{indented}
    \item[] $^*$ Those authors have equally contributed
\end{indented}

\begin{abstract}
\textit{Objective:}
Identifying the spiking activity of alpha motor neurons (MNs) non-invasively is possible by decomposing signals from active muscles, e.g., obtained with surface electromyography (EMG) or ultrasound.
The theoretical background of MN identification using these techniques is convolutive blind source separation (cBSS), in which different algorithms have been developed and validated. 
However, the existence and identifiability of inverse solutions and the corresponding estimation errors are not fully understood. 
In addition, the guidelines for selecting appropriate hyper-parameters are often built on empirical observations, limiting the translation to clinical applications and other modalities.\\
\textit{Approach:}
We revisited the cBSS model for EMG-based MN identification, augmented it with new theoretical insights and derived a framework that can predict the existence of solutions for spike train estimates. 
This framework allows the quantification of source estimation errors due to the imperfect inversion of the motor unit action potentials (MUAP), physiological and non-physiological noise, and the ill-conditioning of the inverse problem. 
To bridge the gap between theory and practice, we used computer simulations.\\ 
\textit{Main results:}
(1) Increasing the similarity of MUAPs or the correlation between spike trains increases the bias for detecting MN spike trains linked with high amplitude MUAPs. 
(2) The optimal objective function depends on the expected spike amplitude, spike amplitude statistics and the amplitude of background spikes. 
(3) There is some wiggle room for MN detection given non-stationary MUAPs, 
(4) There is no connection between MUAP duration and extension factor, in contrast to previous guidelines. 
(5) Source quality metrics like the silhouette score (SIL) or the pulse-to-noise ratio (PNR) are highly correlated with a source's objective function output. 
(6) Considering established source quality measures, SIL is superior to PNR.\\ 
\textit{Significance:}
We expect these findings will guide cBSS algorithm developments tailored for MN identification and translation to clinical applications.
\end{abstract}


\maketitle


\section{Introduction}
Identifying the spiking activity of individual alpha motor neurons (MNs) non-invasively can be achieved by decomposing signals from activated muscles, e.g., obtained with multichannel surface electromyography (EMG) {\citep{Holobar2007a}} or ultrasound {\citep{Rohlen2020}}. 
This indirect motor neuron identification from compound muscle activity is possible because each motor neuron innervates tens to several hundred muscle fibres, amplifying the motor neuron activity on a one-to-one basis \citep{Heckman2012}. 
The motor neuron and all the muscle fibres it innervates are defined as a motor unit (MU).
The activity of each MU can be modelled as a convolution of an impulse response (related to the muscle fibre activity) with the neural discharges of the motor neuron. Thus, template-matching {\citep{DeLuca2006, Nawab2010}} and convolutive blind source separation (BSS) {\citep{Holobar2007b, Negro2016}} are used to blindly estimate the discharges of a (sub)population of motor neurons. 
Although template-matching methods are effective for decomposing intramuscular EMG signals \citep{McGill2005} or under low force levels, these methods face challenges at higher force levels due to increased interference {\citep{Zhou2004}}. 
Convolutive BSS methods are reliable for extracting populations of motor neurons at low to high contraction levels {\citep{Holobar2014,Negro2016,Avrillon2024}}. Although there are (slight) differences between state-of-the-art convolutive BSS-based motor neuron identification algorithms, they are built on the same theoretical background, most importantly, independent component analysis (ICA) {\citep{Hyvarinen2001}}.

Four main steps summarise ICA-based motor neuron identification: 1) reformulate the convolutive mixture model to an instantaneous mixture model, 2) whitening, 3) source estimation by solving an optimisation problem, and 4) post-processing of estimated sources.
The first three steps (extension, whitening, and optimisation) are well described in the seminal Independent Component Analysis (ICA) book published in \citeyear{Hyvarinen2001} by {\citeauthor*{Hyvarinen2001}}. 
The novelty of the EMG-based decomposition methods to generic convolution BSS methods lies in post-processing (fourth step). This includes spike clustering, source quality estimation and (manual) refinement of the predicted spike trains {\citep[e.g.,][]{Negro2016, DelVecchio2020, Avrillon2024b}}.
Although variants of convolutive BSS, e.g., convolution kernel compensation (CKC) {\citep{Holobar2007b,Chen2023,Lin2024,Xia2024,Chen2025}} and fastICA {\citep{Chen2015,Negro2016,Lundsberg2023,Chen2024}}, are widely used and validated for the decomposition of surface EMG or ultrasound data, the selected hyper-parameters of the algorithms are often built on empirical observations rather than theory.
Since surface EMG has been the first and still is the most important experimental modality for the non-invasive detection of motor neuron spike trains, many decomposition methods have been tailored for EMG signals. 
Yet, empirical observations, e.g., from the decomposition of ultrasound-based image sequences {\citep{Lubel2024, Rohlen2025}}, challenge some assumptions and guidelines of algorithms used for motor neuron identification.

A fundamental requirement for the application of ICA is mutually independent sources, which motor neuron spike trains violate due to having delayed sources in the extension step {\citep{Farina2016a}} and common synaptic input(s). 
While ICA is known to be robust against modest violations of the independence assumption \citep{Hyvarinen2000}, the independence violation has also been explicitly acknowledged for surface EMG decomposition {\citep{Farina2016a}}.
It has been argued that due to the sparsity of the MU spike trains, the sum of spike trains is more Gaussian than a single spike train, making non-Gaussianity (a common optimization goal in ICA) also a tractable choice for separating dependent but sparse sources.
\textit{
Nevertheless, the role of latent source dependencies is not fully understood in convolutive BSS-based motor neuron identification methods.}

The decomposition problem is typically ill-conditioned. 
This means that the mixing matrix is not invertible, and even if the forward model is known, it is impossible to recover the activity of a complete motor neuron pool {\citep[e.g.,][]{Klotz2023}}.
Typically, low amplitude MUAPs (either due to size or depth) and MUs with non-unique MUAPs are considered non-detectable based on empirical rules {\citep[e.g.,][]{Farina2008}}.
Further, the effect of noise has been studied mostly empirically or, if investigated theoretically, under a simplified Gaussian white noise assumption {\citep[e.g.,][]{Holobar2007a, Negro2016, Lubel2024}}.
\textit{
Thus, the existence and identifiability of inverse solutions and the corresponding estimation errors are not fully understood.} 

Most EMG-based decomposition methods assume the mixing matrix is stationary {\citep{Negro2016}}. 
For surface EMG, this approximately holds for isometric and non-fatiguing contractions. 
However, it has been shown that convolutive BSS methods with moderate adjustments can handle non-stationarities, e.g., given slow \citep{Oliveira2021, Guerra2024} or cyclic motions \citep{Glaser2018}.
Moreover, the ultrasound-based motor neuron identification using linear convolutive BSS works reliably {\citep{Rohlen2025}}, although the MU twitches cannot be considered stationary {\citep{Raikova2007}}.
\textit{
Therefore, the limits of the standard linear convolution BSS model in the presence of non-stationarities must be systematically characterised.}

Guidelines for EMG-based decomposition suggest that the impulse response duration must be shorter than the inter-spike interval {\citep{Chen2015}}. 
Although the mechanical impulse response after a neural discharge (MU twitch) is approximately 10 times longer than the electrical impulse response (MUAP) and at least two to three times larger than typical inter-spike interval durations {\citep{McNulty2000, Raikova2008}}, a convolutive BSS method applied to sequences of ultrasound images could still identify motor neuron spike trains \citep{Lubel2024}. 
\textit{This example demonstrates that the guidelines for selecting appropriate hyperparameters must be refined.}

The overall purpose of the study was to update the understanding of the EMG-based convolutive BSS method for motor neuron identification. 
We aimed to summarize state-of-the-art convolution BSS methods used for motor neuron identification and derive new theoretical tools that allow studying the existence and identifiability of physiological solutions predicted by convolutive BSS models.
The theory is underlined by computer simulations combining a realistic motor neuron pool model and a database of MUAPs extracted from experimental recordings.
In this way, we intend to bridge the gap from theory to application, which is important for translating motor neuron identification methods to other signal modalities like ultrasound and clinical applications.

We found that increasing the similarity of MUAPs or the correlation between spike trains increases the bias for detecting MN spike trains linked with high-amplitude MUAPs. Moreover, there is no connection between MUAP duration and extension factor, and there is some wiggle room for MN detection given non-stationary MUAPs. Robust learning and the optimal objective function depend on the expected spike amplitude, spike amplitude statistics, and the background spike amplitudes. Finally, source quality metrics are highly correlated with a source’s objective function output.

\section{Theory}

\subsection{The convolutive mixture model}
A multichannel EMG signal can be described as a linear convolution mixture, i.e.,
\begin{equation}
\label{eqn:convBSSspiketrain}
x_i(t) \ = \ \sum_{j=1}^{N} \sum_{l=0}^{L-1}h_{ij}(l) s_j(t-l) \, + \, \varepsilon_{i} (t) \ ,
\end{equation}
where $x_i(t)$ (with $i=1,..., M$ and $t=0,..., T$) denotes the $i$th channel of the EMG signal with $M$ electrodes, and $t$ denotes a discrete time frame with $T+1$ being the total number of samples. 
The impulse response of the $j$th MU, i.e., the MUAP, at the $i$th channel is given by $h_{ij}(l)$, where $L$ denotes the number of discrete samples of the impulse responses and $N$ is the number of active MUs in the EMG signal. 
Further, $\varepsilon_{i}$ denotes additive noise.
The spike train for the $j$th MU is defined as
\begin{equation}
\label{eqn:convBSSdirac}
s_j(t)=\sum_{r \in \mathcal{S}_j} \delta (t-t_j^r),
\end{equation}
where $\mathcal{S}_j:=\{t_j^1, ..., t_j^{T_j}\}$ is the set of discharge times of the $j$th MU and $\delta(\cdot)$ is the Dirac delta function. 

A convolutive mixture with finite impulse response filters can be represented as a linear instantaneous mixture in the time domain. 
This is achieved by defining an extended vector of sources and observations, each including the original sources or observations and their delayed versions {\citep{Hyvarinen2001}}. 

This yields
\begin{equation}
\label{eqn:convBSSextendedmatrix}
\begin{split}
\mathbf{\tilde{x}}(t) \ &= \ \mathbf{\tilde{H}} \mathbf{\tilde{s}}(t) \, + \, \bm{\tilde{\varepsilon}}(t) \ , \  \text{with} \\[3mm]
\mathbf{\tilde{H}} \ &= \ 
\begin{bmatrix}
\mathbf{\tilde{H}}_{11} & \cdots & \mathbf{\tilde{H}}_{1N} \\
\vdots & \ddots & \vdots \\
\mathbf{\tilde{H}}_{M1} & \cdots & \mathbf{\tilde{H}}_{MN} \\
\end{bmatrix}, \\[2mm]
\mathbf{\tilde{x}}(t) \ &= \ [\mathbf{\tilde{x}}_1(t),\mathbf{\tilde{x}}_2(t),...,\mathbf{\tilde{x}}_M(t)]^T \ ,\\[1mm]
\mathbf{\tilde{s}}(t) \ &= \ [\mathbf{\tilde{s}}_1(t),\mathbf{\tilde{s}}_2(t),...,\mathbf{\tilde{s}}_N(t)]^T \ ,\\[1mm]
\bm{\tilde{\varepsilon}}(t) \ &= \ [\bm{\tilde{\varepsilon}}_1(t),\bm{\tilde{\varepsilon}}_2(t),...,\bm{\tilde{\varepsilon}}_M(t)]^T \ ,\\
\end{split}
\end{equation}
where the block matrices or vectors are given by
\begin{equation}
\label{eqn:convBSSextendedcomps}
\begin{split}
\mathbf{\tilde{H}}_{ij} \ &= \ 
\begin{bmatrix}
h_{ij}(0) & \cdots & h_{ij}(L-1) & 0 & \cdots & 0 \\
0 & \ddots & \ddots & \ddots & \ddots & \vdots \\
\vdots & \ddots & \ddots & \ddots & \ddots & 0 \\
0 & \cdots & 0 & h_{ij}(0) & \cdots & h_{ij}(L-1) \\
\end{bmatrix} \ , \\[2mm]
\mathbf{\tilde{x}}_i(t)&=[x_i(t),x_i(t-1),\ldots,x_i(t-R)] \ ,\\[1mm]
\mathbf{\tilde{s}}_j(t)&=[s_j(t),s_j(t-1),\ldots,s_j(t-L-R+1)] \ ,\\[1mm]
\bm{\tilde{\varepsilon}}_i(t)&=[\varepsilon_i(t),\varepsilon_i(t-1),\ldots,\varepsilon_i(t-R)] \ .\\
\end{split}
\end{equation}

The mixing model in Equation~\eqref{eqn:convBSSextendedcomps} is typically ill-conditioned. Thus, the activity of a full motor neuron pool cannot be reconstructed even when the forward mixing model is known {\citep[e.g.,][]{Klotz2023}}. 
Treating non-detectable sources as part of the additive noise {\citep{Negro2016, Farina2016a}}, we get:
\begin{equation}
\label{eqn:convBSSnondetect}
\begin{split}
\mathbf{\tilde{x}}(t)\ &= \ \mathbf{\tilde{H}}_{:,j
_\mathcal{D}} \, \mathbf{\tilde{s}}_{j_\mathcal{D}}(t) \, + \, \mathbf{\tilde{H}}_{:,j_\mathcal{N}} \mathbf{\tilde{s}}_{j_\mathcal{N}}(t) \, + \,\bm{\tilde{\varepsilon}}(t) \ , \\
&= \ \mathbf{\tilde{H}}_{:,j_\mathcal{D}} \, \mathbf{\tilde{s}}_{j_\mathcal{D}}(t) \, + \,\bm{\tilde{\varepsilon}}'(t) \ .
\end{split}
\end{equation}
Therein, $\mathbf{\tilde{H}}_{:,j}$ is the $j$th column of block matrix $\mathbf{\tilde{H}}$ and $\mathbf{\tilde{s}}_{j}$ is the $j$th row of the block vector $\mathbf{\tilde{s}}$.
Further, $\mathcal{D} := \{j_\mathcal{D}^1, ..., j_\mathcal{D}^{N_\mathcal{D}}\}$ is the set of detectable sources, with column indices $j_\mathcal{D}$, and $\mathcal{N} := \{j_\mathcal{N}^1, ..., j_\mathcal{N}^{N_\mathcal{N}}\}$ is the set of non-detectable sources, with column indices $j_\mathcal{N}$.
The total number of sources (i.e., the MN spike trains and their delayed versions) is $R\cdot N=N_\mathcal{D}+N_\mathcal{N}$, with $N_\mathcal{D}$ or $N_\mathcal{N}$ denoting the number of detectable and non-detectable sources, respectively.

\subsection{The inverse problem}
The goal of decomposing EMG signals is to estimate the (detectable) motor neuron spike trains only given the extended observation matrix. 
Most state-of-the-art convolutive BSS-based decomposition algorithms can be classified as variants of ICA.
In this section, we revisit the fundamentals of ICA to derive new theoretical insights into the performance of motor neuron identification algorithms.

\subsubsection{Preconditioning}\label{sec:inverse_preconditioning}
In the attempt to solve the inverse problem, it is common to conduct a few pre-processing steps.  
This includes, if possible, rejecting parts of the noise, e.g., using bandpass or notch filters. 
Due to the linearity of the mixing model (cf. Equation~\ref{eqn:convBSSextendedmatrix}), only linear pre-processing steps should be conducted {\citep{Hyvarinen2001}}.
Next, the extended observations are centred by subtracting the (sample) mean from the extended observations.   
Then, a whitening transformation is applied 
\begin{equation}
\label{eqn:convBSSwhite}
    \mathbf{\tilde{z}}(t) \ = \ \mathbf{V} \mathbf{\tilde{x}}(t)  \ ,
\end{equation}
where the whitening matrix $\mathbf{V}$ is constructed such that the covariance matrix of extended and whitened observations $\mathbf{\tilde{z}}(t)$ is the identity matrix, i.e., $\mathbf{C}_{\tilde{z}\tilde{z}} = \mathbf{I}$.

Note that the whitening transformation is always possible; however, it is not unique. Popular choices are principal component analysis (PCA) or zero-phase component analysis (ZCA) whitening, where the whitening matrix is constructed from the eigendecomposition of the extended signal's covariance matrix {\citep{Krizhevsky2009}}.
Practically, one needs to empirically estimate the covariance matrix given the extended observations. It turns out useful to decompose the covariance matrix of the extended signal in Equation~\eqref{eqn:convBSSnondetect} into {\citep[e.g.,][]{Hyvarinen2001}}
\begin{equation}\label{eqn:covariance_decomposition}
    \mathbf{C}_{\mathbf{\tilde{x}}\mathbf{\tilde{x}}} \ = \ \mathbb{E}\left[\mathbf{\tilde{x}}(t) \mathbf{\tilde{x}}^T(t)\right] \ = \ \mathbf{\tilde{H}}_{:,j
_\mathcal{D}}  \mathbf{C}_{\mathbf{\tilde{s}}_{j_\mathcal{D}}\mathbf{\tilde{s}}_{j_\mathcal{D}}}\mathbf{\tilde{H}}_{:,j
_\mathcal{D}}^T \, + \, \mathbf{C}_{\mathbf{\tilde{\varepsilon}'}\mathbf{\tilde{\varepsilon}'}} \ .
\end{equation}
Therein, $\mathbf{C}_{\mathbf{\tilde{x}}\mathbf{\tilde{x}}}$ denotes the covariance matrix of the extended observations, $\mathbf{C}_{\mathbf{\tilde{s}}_{j_\mathcal{D}}\mathbf{\tilde{s}}_{j_\mathcal{D}}}$ is the covariance matrix of the extended detectable sources and $\mathbf{C}_{\mathbf{\tilde{\varepsilon}'}\mathbf{\tilde{\varepsilon}'}}$ is the covariance matrix of the extended noise including the non-detectable MUs.
Further, $\mathbb{E}(\cdot)$ denotes the expectation operator.
The whitening transformation attempts to orthogonalise the mixing matrix and already solves fifty percent of the BSS problem {\citep{Hyvarinen2000}}. 
Moreover, as the inverse of an orthogonal matrix is its transpose, one can construct the inverse mixing matrix column-by-column. 
Thus, the $k$th spike train is then reconstructed by applying a suitable projection vector $\mathbf{w}_k^T$, i.e., an estimate of one column of the mixing matrix, to the extended and whitened observations:
\begin{equation}
\label{eqn:convBSSwk}
    \widehat{\tilde{s}}_{k}(t) \ = \ \mathbf{w}_k^T \mathbf{\tilde{z}}(t) \ .
\end{equation}
Therein $\widehat{\tilde{s}}_{k}(t)$ is the estimate of the $k$th detectable spike train, which is a non-unique solution for the inverse of the linear system given in Equation~\eqref{eqn:convBSSextendedmatrix} (e.g., applying a linear scaling factor is always possible). 
Thus, we will require $\mathbf{w}_k$ as a unit vector for the following investigations.
In the literature, the spike estimation is sometimes conducted in the non-whitened space {\citep[e.g.,][]{Holobar2007a, Farina2016a}}.
It can be shown (see \ref{appendix:ckc_spike_estimation}) that if the whitening matrix is invertible, this approach is equivalent to Equation~\eqref{eqn:convBSSwk}. 

\subsubsection{Existence of an inverse solution}\label{sec:existence}
Inserting Equations~\eqref{eqn:convBSSnondetect} and \eqref{eqn:convBSSwhite} into Equation~\eqref{eqn:convBSSwk}, the estimate of the $k$th spike train (with delay $l$) can be expressed in terms of a dot product. In detail, the projection vector is applied to each column of the extended and whitened mixing matrix weighted by the corresponding source activity (i.e., a sequence of ones and zeros) and the noise. 
We obtain
\begin{equation}
\label{eqn:convBSSallerrors}
\begin{split}
\widehat{\tilde{s}}_{k}(t) \ &=  \ \mathbf{w}_k^T \left(\mathbf{V} \mathbf{\tilde{H}}_{:,j
_\mathcal{D}} \, \mathbf{\tilde{s}}_{j_\mathcal{D}}(t) \, + \mathbf{V} \mathbf{\tilde{H}}_{:,j
_\mathcal{N}} \, \mathbf{\tilde{s}}_{j_\mathcal{N}}(t) \, + \, \mathbf{V} \bm{\tilde{\varepsilon}}(t)  \right) \\
&= \ \Big\langle \mathbf{w}_k \ , \  \tilde{s}_{kl}(t) \mathbf{\tilde{h}}_{kl} \, + \!\!\! \sum_{(j,l) \in \mathcal{M}} \!\! \tilde{s}_{jl}(t) \mathbf{\tilde{h}}_{jl} \, + \, \bm{\tilde{\varepsilon}}^\mathrm{w}(t)   \Big\rangle \\
&= \ \widehat{\tilde{s}}{}_k^{(k,l)}(t) \, + \!\!\! \sum_{(j,l) \in \mathcal{M}} \!\! \widehat{\tilde{s}}{}_{k}^{(j,l)}(t) \, + \, \varepsilon_k(t) \ ,
\end{split}
\end{equation}
where $\mathbf{\tilde{h}}_{ul}$ are the columns of the extended and whitened mixing matrix (see Equation~\eqref{eqn:MUAP_filter}), $\tilde{s}_{ul}(t)$ denotes the $l$th delayed spike train of MU $u$ (with $u$ denoting an arbitrary MU) and the set $\mathcal{M}$ includes all sources with indices $(j,l)\neq (k,l)$.
Further, $\widehat{\tilde{s}}{}_k^{(k,l)}(t)$ is the contribution of MU $k$ (with delay $l$) to the estimated spike strain, $\widehat{\tilde{s}}{}_k^{(j,l)}(t)$ are background spikes due to other sources and their delayed versions, and $\varepsilon_k(t)$ is the projected noise. 

Finding an inverse solution for the spike train of MU $k$ with delay $l$ (i.e., a reconstruction of $\tilde{s}_{kl}(t)$) requires the existence of a projection vector $\mathbf{w}_k$ such that the term $\widehat{\tilde{s}}{}_k^{(k,l)}(t)$ in Equation~\eqref{eqn:convBSSallerrors} dominates. 
When the sources are (sufficiently) uncorrelated, the expected value of the reconstructed spike train $\widehat{\tilde{s}}_{k}(t)$ when a source is activated (i.e., $\tilde{s}_{ul}(t)=1$) can be calculated by the dot product between the projection vector $\mathbf{w}_k$ and the columns of the whitened and extended mixing matrix $\mathbf{\tilde{h}}_{ul}$.
Given that the projection vector $\mathbf{w}_k$ is a unit vector we obtain
\begin{equation}
\label{eqn:source_amplitude_estimate}
 \mathbb{E}\Big(\widehat{\tilde{s}}_k(t \in \mathcal{S}_{u})\Big) \ = \ \langle \mathbf{w}_k, \mathbf{\tilde{h}}_{ul}\rangle \ = \ S_\mathrm{cos}^{k,ul} \cdot ||\mathbf{\tilde{h}}_{ul}|| \ = \ \rho_k^{(u,l)} \ , 
\end{equation}
where $\mathcal{S}_{u}:=\{t_{u}^1,...,t_{u}^{T_{u}}\}$ is the set of spike times of the $u$th source with delay $l$, $S_\mathrm{cos}^{k,ul}$ is the cosine similarity between $\mathbf{w}_k$ and $\mathbf{\tilde{h}}_{ul}$, and $||\cdot||$ denoting the norm of an arbitrary vector. Thereby note that $\mathbf{\tilde{h}}_{ul}$ (see Equation~\eqref{eqn:MUAP_filter}) does not purely reflect the MU impulse response as the sources' statistics that are included through the whitening transformation (see Equation~\eqref{eqn:covariance_decomposition}). Further note, the cosine similarity between two vectors is one if the vectors are aligned (i.e., parallel) and zero if the vectors are orthogonal.
Thus, the amplitude of the spikes of interest becomes maximal if $\mathbf{w}_k = \mathbf{\tilde{h}}_{kl} \, / \, ||\mathbf{\tilde{h}}_{kl}||$ (i.e., $S_\mathrm{cos}^{k,kl}=1 $) and $l$ is chosen such that the norm of the selected column is maximal (i.e., $\operatorname{arg \ max} ||\mathbf{\tilde{h}}_{kl}||$, with $0<l<L-1$).
Analogous, the expected amplitude of background spikes depends on the cosine similarity between $\mathbf{w}_k$ and all other columns of the extended and whitened mixing matrix $\mathbf{\tilde{h}}_{jl}$ with $(j,l) \in \mathcal{M}$. In an average sense, one can define the separability (SEP) of a MU $k$ as the relative peak separation between the expected amplitudes of the true spikes and the largest background spikes, i.e.,
\begin{equation}
\label{eqn:separation_of_MUs}
    \text{SEP}_k \ = \ 1 \, - \, \frac{1}{S_\mathrm{cos}^{k,kl} \cdot ||\mathbf{\tilde{h}}_{kl}||} \ \underset{(j,l)\in \mathcal{M}}{\operatorname{arg \, max}}\left(S_\mathrm{cos}^{k,jl} \cdot ||\mathbf{\tilde{h}}_{jl}||\right) \ .
\end{equation}
A peak separation larger than zero is a necessary condition for the identifiability of an MU. 
However, spikes can be overlayed by noise, i.e., the activity of other non-detectable MUs or any other noise source summarized by $\bm{\tilde{\varepsilon}}'(t)$ in Equation~\eqref{eqn:convBSSnondetect}. 
To showcase this, we compute the variance of the projected noise, i.e.,  
\begin{equation}
\label{eqn:projectedNoise}
\begin{split}
     \sigma_\mathrm{n}^2 \ = \ \operatorname{Var}\Big(\big\langle \mathbf{w}_k  ,  \mathbf{\tilde{z}}^\mathrm{n}(t)   \big\rangle\Big) &= \ \sum_{i=1}^{M\cdot R} w_{k,i}^2 \operatorname{Var}\Big(\tilde{z}_i^\mathrm{n}(t)\Big) \ = \ \sum_{i=1}^{M\cdot R} w_{k,i}^2 \sigma_i^2 \ , \ \text{with} \\
     \mathbf{\tilde{z}}^\mathrm{n}(t) \ &= \!\! \sum_{(j,l)\in \mathcal{M}} \!\! \tilde{s}_{jl}(t) \mathbf{\tilde{h}}_{jl} \, + \, \bm{\tilde{\varepsilon}}^\mathrm{w}(t),
\end{split}    
\end{equation}
where $\mathbf{\tilde{z}}^\mathrm{n}(t)$ denote the whitened noise terms in Equation~\eqref{eqn:convBSSallerrors}. It can be seen that the influence of the noise becomes zero if the projection vector is orthogonal to the projected noise, i.e., $\langle \mathbf{w}_k  ,  \mathbf{\tilde{z}}^\mathrm{n}(t)\rangle = 0$. As an upper bound, one can assume $\mathbf{\tilde{z}}^\mathrm{n}(t) \approx \mathbf{\tilde{z}}(t)$, which has unit variance by construction.
The robust identification of a MU $k$ requires that a projection vector $\mathbf{w}_k$ exist such that
\begin{equation}
\label{eqn:identifyMUs}
    \text{SEP}_k \cdot S_\mathrm{cos}^{k,kl} \cdot ||\mathbf{\tilde{h}}_{kl}||  \  >   \ \lambda \cdot \sigma_\mathrm{n} \ .
\end{equation}
Therein, $\lambda$ is the desired uncertainty bound describing how many standard deviations the expectation of the peaks of interest is above the highest background peak.

\subsubsection{Identifiability of an inverse solution}\label{sec:identifiability}
The blind identification of the (delayed) sources in Equation~\eqref{eqn:convBSSnondetect} is a linear BSS problem {\citep{Hyvarinen2001}}.
The fundamental assumptions for using ICA to solve a BSS problem are
1) linear superposition of the observations, 2) the prior of the sources, and 3) the joint prior of the sources being factorial (i.e., statistical independence).
When the sources are uncorrelated (approximately achieved through the whitening transformation, cf. Section~\ref{sec:inverse_preconditioning}), non-Gaussian sources are sufficient {\citep{Hyvarinen1998a}}.

One can thus blindly estimate the inverse mixing matrix by searching for projection vectors $\mathbf{w}_k$ (with $||\mathbf{w}_k||=1$) predicting maximally non-Gaussian sources.
In detail, we define an objective function 
\begin{equation}\label{eqn:def_loss_function}
\begin{split}
    L(\mathbf{w}_k) \ &= \ \sum_t \mathbb{E} \Big(G(\mathbf{w}_k^T \mathbf{\tilde{z}}(t)) \Big)  \ = \ \sum_t \mathbb{E} \Big(G(\widehat{\tilde{s}}_{k}(t) \Big) \ ,
\end{split}    
\end{equation}
where $G(\cdot)$ is an arbitrary non-linear and non-quadratic function. 
A popular choice is statistical moments of order higher than two (e.g., skewness or kurtosis).
Finding an optimum of Equation~\eqref{eqn:def_loss_function} requires an optimisation algorithm, e.g., the (natural) gradient descent or an approximated Newton solver (as done in fastICA).\\

To study the given optimisation problem, we split up the sum term in Equation~\eqref{eqn:def_loss_function} into the spikes of an MU of interest (with delay $l$), another MU highlighted by the separation vector $\mathbf{w}_k$, and the projected noise. Assuming that the expected amplitude at the spikes is $\rho_k^{(u,l)} = S_\mathrm{cos}^{k,ul} ||\mathbf{\tilde{h}}_{ul}||$ (see Equation~\ref{eqn:source_amplitude_estimate}) and that the expected value of the noise is zero, the expectation in each term is approximated by a Taylor series (see \ref{appendix:taylor_series_cf}) to obtain
\begin{equation}
\label{eqn:approx_loss}
\begin{split}
    L(\mathbf{w}_k) \ & = \ \sum_{t\in\mathcal{S}_k} \mathbb{E}\Big[ G\Big(\widehat{\tilde{s}}_{k}(t)  \Big)\Big] \, + \,  \sum_{t\in\mathcal{S}_j} \mathbb{E}\Big[ G\Big(\widehat{\tilde{s}}_{k}(t)   \Big)\Big] \, + \  \sum_{t\in\mathcal{S_N}} \mathbb{E}\Big[ G\Big(\widehat{\tilde{s}}_{k}(t)  \Big)\Big]  \\ 
    &= \ T_k \Bigg( G\left(\rho_k^{(k,l)}\right)  +    G''\left(\rho_k^{(k,l)}\right) \frac{\sigma_k^2}{2} \, +   G^{(3)}\left(\rho_k^{(k,l)}\right) \frac{\gamma_1^k\sigma_k^3}{6} \, + \, ... \Bigg) \\
    &\, + \,  T_j \Bigg( G\left(\rho_k^{(j,l)}\right) +  G''\left(\rho_k^{(j,l)}\right) \, \frac{\sigma_j^2}{2} +  G^{(3)}\left(\rho_k^{(j,l)}\right) \frac{\gamma_1^j\sigma_j^3}{6} \, + \, ... \Bigg) \\
    &\, + \, T_{\mathcal{N}} \Bigg( G(0) +   G''(0)  \frac{\sigma_\mathcal{N}^2}{2} +  G^{(3)}\left(0\right)\frac{\gamma_1^\mathcal{N}\sigma_\mathcal{N}^3}{6}  \, + \, ... \Bigg) \, + \, \mathcal{O}(s^4)  \ .
\end{split}    
\end{equation}
where $\sigma$ denotes the standard deviation and $\gamma_1$ denotes the skewness. 
Moreover, $\mathcal{S}_k := \{{t^1_k, ...,t_k^{T_k}}\}$ are the spikes of MU $k$ (with delay $l$) and $\mathcal{S}_j := \{t_j^1, ...,t_j^{T_j}\}$ are the spikes of MU $j$ (with delay $l$).
Lastly, $\mathcal{S_N}:= \{{t^1_\mathcal{N}, ...,t_\mathcal{N}^{T_\mathcal{N}}}\}$ denote all time points associated with noise. 

It can be seen that a local optimum of $L(\mathbf{w}_k)$ is equivalent to learning a column of the extended and whitened mixing matrix if the zero-order term associated with MU $k$ (with delay $l$) dominates, i.e., $\mathbf{w}_k \xrightarrow{} \alpha \,  \mathbf{\tilde{h}}_{kl}$, where $\alpha$ is an arbitrary scaling parameter. 
However, (local) maxima of the objective function can potentially be associated with maximizing any other term in Equation~\eqref{eqn:approx_loss}. 
This means that physiologically spurious projection vectors $\mathbf{w}_k$ are part of the mathematical solution space, making interpreting decomposition results challenging.
For example, the objective function will further increase if at least one additional MU is sensitive to $\mathbf{w}_k$. 
A projection vector associated with a maximum can thus become a mixture of two (or more) sources.
Further, the objective function rewards noisy spike trains and outlier spikes due to the higher-order terms in Equation~\eqref{eqn:approx_loss}, which can cause overfitting to these time instances.

\begin{figure}[ht!]
    \includegraphics[width=\textwidth]{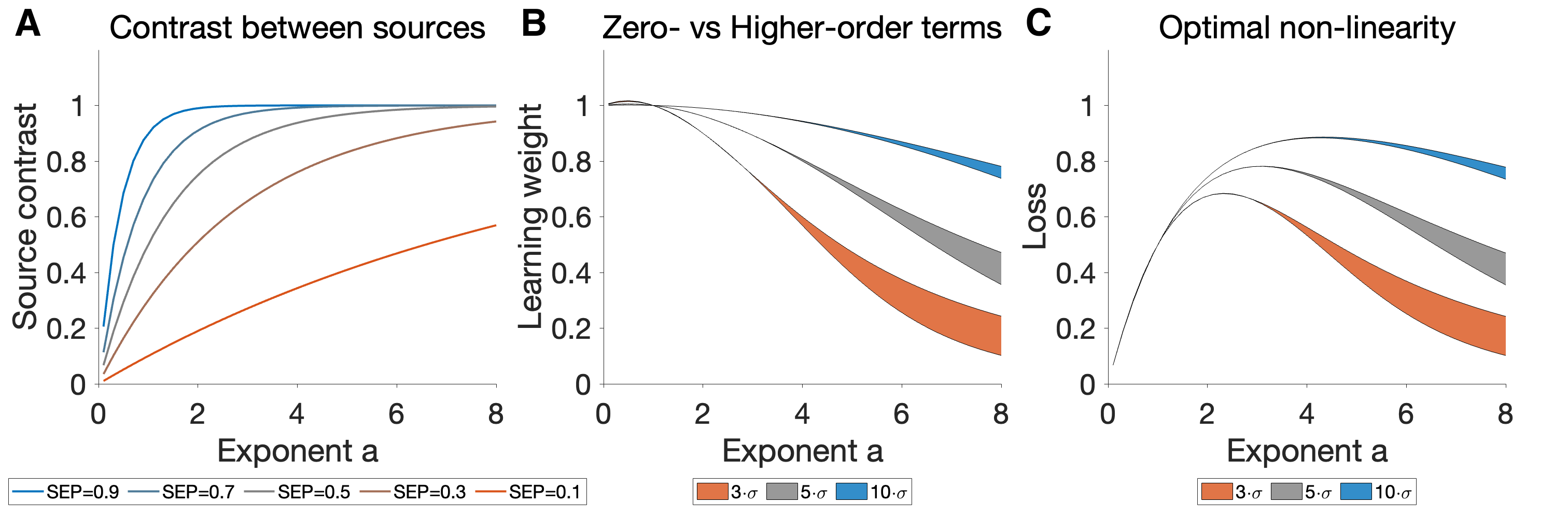}
    \caption{(\textbf{A}) Source contrast between two motor units (MUs) after applying a non-linear function $G(s)$. The colours show different relative peak separation values (SEP). (\textbf{B}) Learning goodness is quantified as the ratio between the zero-order term and the overall contribution of the peak cluster to the objective function. The colours show different spike amplitudes (relative to the standard deviation $\sigma$). The upper bound is up to the second-order term; the lower bound is up to the fourth-order term. (\textbf{C}) The optimal non-linearity corresponds to the maximum of the product between the source contrast and the learning weight. Here, we assumed $\text{SEP}=0.7$.}
    \label{fig:effect_of_exponent}
\end{figure}

A suitable choice of non-linearity can maximize the contrast between MUs with similar extended and whitened MUAPs and limit the influence of overfitting.
To illustrate the influence of the selected non-linearity $G(\cdot)$, we will investigate the class of arbitrary power functions
\begin{equation}\label{eqn:power_law_contrast_fcn}
    G(s) \ = \ \operatorname{sgn}(s) \cdot |s|^a \ , 
\end{equation}
where the exponent $a \in \mathbb{R}$ is an adjustable parameter and $\operatorname{sgn}(s)$ denotes the signum function.
We showcase the influence of the objective function on the identifiability of MUs through a series of examples. 
First, we consider two sources and vary the relative peak separation (SEP) from 0.1 to 0.9 (Figure~\ref{fig:effect_of_exponent}A). 
It can be observed that in the case where the relative peak separation is close to one, the specific choice of $a$ is not critical as the source contrast, i.e., the normalized difference between the non-linearity applied to the spike amplitude and the non-linearity applied to the background spike amplitude, quickly approaches one.
In the challenging case where SEP approaches zero, a suitable choice of $G(\cdot)$ can be essential to provide sufficient contrast between the spikes of different MUs.  
Here, increasing the exponent $a$ improves the separation (Figure~\ref{fig:effect_of_exponent}A).

Practically, the degree of the selected non-linearity is bounded by the higher-order terms (i.e., noise or outlier spikes). 
This is illustrated by considering a spike train with 300 discharges with an amplitude of 5, 10, or 20 times the spike amplitude's standard deviation (i.e., spanning the typical range for real decompositions).  
To consider the effect of higher-order terms (i.e., larger than two), we assume that there exists one outlier spike with an amplitude of \SI{150}{\%} of the mean spike amplitude.
Figure~\ref{fig:effect_of_exponent}B shows the learning weight, i.e., the relative contribution of the zero-order term of the peak cluster to the overall value of the objective function, depending on the variability of the spike amplitudes and the presence of outlier spikes. 
The upper bound only considers the effect of the variance of the spike amplitudes (i.e., the second-order term). The lower bound values consider terms up to an order of four.
Robust decomposition requires a learning weight close to one.
It is observed that the decay of the learning weight depends on the amplitude, the variability, and higher-order statistical moments of the spikes.

As a criterion for the optimal exponent, we consider the product between the source contrast and the learning weight.
In Figure~\ref{fig:effect_of_exponent}C, this is illustrated by considering a relative peak separation of 0.7 (typical values range from 0.5 to 0.8). 
It is observed that the optimal non-linearity is a compromise between spike contrast and limiting errors due to overfitting.
Thereby, the optimal exponent $a$ increases when the normalized spike amplitude (relative to the spike variance) increases and the relative peak separation decreases.
For example, assuming $\text{SEP}=0.7$ the optional exponents are 2.31 (amplitude: $5\,\sigma$), 3.12 (amplitude: $10\,\sigma$) or 3.92 (amplitude: $20\,\sigma$) and increase to 3.12, 4,32 and 5.53 respectively given $\text{SEP}=0.5$.
Lastly, we note that a typical non-physiological solution in ICA-based source estimates is a single spike. \ref{appendix:single_spike} provides a condition on how single spike solutions are related to the selected non-linearity $G(s)$.
 
\subsection{Spike clustering and source quality metrics}\label{sec:source_quality}
The final step in the decomposition is extracting the motor neuron discharge times from the reconstructed sources. 
Although not strictly necessary, spike clustering is often achieved after applying a non-linear function $F(s)$, similar to the objective function given in Equation~\eqref{eqn:def_loss_function}, to the $k$th reconstructed spike train:
\begin{equation}
    \varphi^k(t) = F(\widehat{\tilde{s}}_{k}(t)) \ . 
\end{equation}
A common choice is the asymmetrical power law given in Equation~\eqref{eqn:power_law_contrast_fcn} with an exponent of $a=2$. 
Typically, first, a peak finding algorithm is applied to $\varphi^k(t)$, followed by a clustering algorithm (e.g., K-means) aiming to separate the spikes associated with the activity of MU $k$ from (potential) false positive spikes.

A fundamental challenge is quantifying the goodness of the extracted binary spike train without knowing the ground truth.
The most prominent examples of such source quality metrics are the silhouette (SIL) score {\citep{Negro2016}} and the pulse-to-noise ratio (PNR) {\citep{Holobar2014}}.
Often, a SIL (see Equation~\eqref{eqn:sil_alt}) value larger than $0.9$ is considered an indicator of robust source identification \citep{Negro2016}. 
It can be shown (see \ref{appendix:source_quality_metrics}) that a SIL value of $0.9$ approximately corresponds to the case where the peak and noise clusters are separated by $3$ standard deviations of $\varphi(t_i)$.
However, it does not reflect the variability of the background spikes and the effect of higher-order statistics, limiting its sensitivity to false positive and negative rates, particularly in the presence of outlier spikes. 
Considering PNR (see Equation~\eqref{eqn:def_pnr}), a threshold of \SI{30}{dB} is often used as an indicator of robust MU source extraction.  
Using $\varphi^k(t) = \operatorname{sgn}(\widehat{\tilde{s}}_{k}(t))\cdot \widehat{\tilde{s}}_{k}(t)^2$, it is obvious that PNR is the ratio of the contribution of the spike cluster and the noise cluster (background peaks and noise) given a $4$th order objective function (i.e., kurtosis). 
However, due to the contribution of higher order terms in the objective function (see Equation~\eqref{eqn:approx_loss}), PNR also provides limited insights into the false negative and the false positive rate (see {Figure~\ref{fig:Separabilitymetric}B-C}).

\section{In silico experiments}
We performed in silico experiments to study the convolutive BSS-based MU identification systematically. Motor neuron spike trains are simulated using a leaky integrate-and-fire model. EMG signals are simulated by convolving the motor neuron spike trains with experimentally extracted MUAPs.

\subsection{Motor neuron pool model}
We considered a leaky integrate-and-fire model to model a pool of motor neurons. 
This approach offers an attractive trade-off between simplicity and capturing many important neuronal characteristics. 
The evolution of the $j$th motor neuron's membrane potential $V_j(t)$ at time $t$ is given by 
\begin{equation}
\label{eqn:LIF}
\tau_j \frac{d V_j(t)}{d t} =
\begin{cases}
   &0 \qquad \qquad \qquad \qquad \qquad \qquad   , \ \text{for} \ \ t \in [t^\mathrm{s}_j, t^\mathrm{s}_j+t_j^\mathrm{ref}] \\
   &-g_{j}^\mathrm{L}\Big(V_j(t)-V_\mathrm{r}\Big) \, + \, 
    g_{j}^\mathrm{E}R_j I_j(t) \ , \ \text{otherwise}.  \\
\end{cases}    
\end{equation}
With the additional rules, that (i) a spike $t_j^\mathrm{s}$ of the $j$th motor neuron is recorded when $V_j(t) \geq V_\mathrm{th}$, with $V_\mathrm{th}$ denoting the threshold potential, and (ii) at each spike the membrane potential is reset to the resting potential $V_\mathrm{r}$ for a refractory period $t_{j}^\mathrm{ref}$, i.e., where the neuron cannot depolarize, which depend on the size of the motor neuron.
We used the initial condition $V_j(0)=V_\mathrm{r}$.
Further, $R_j$ is the membrane resistance for the $j$th motor neuron, $I_j(t)$ is the synaptic input for the $j$th motor neuron at time $t$, $\tau_j$ is a membrane time constant for the $j$th motor neuron. 
Finally, $g_{j}^\mathrm{L}$ and $g_{j}^\mathrm{E}$ are parameters to modulate the time constant for the leakage and the excitability of the membrane. The evolution of the membrane potential (i.e., Equation~\ref{eqn:LIF}, excluding the refractory times) was computed using an explicit Euler method. 
The model parameters (see \ref{appendix:mn_params}) were extracted from experimental animal models {\citep{Caillet2022}}. 

\subsection{Synaptic input}
The synaptic input $I_j(t)$ to each motor neuron comprised common excitatory input, common noise, and independent noise \citep[e.g.,][]{Negro2011, Rohrle2019}. 
The common excitatory drive was prescribed as a trapezoidal signal (\SI{10}{s} ramping up, \SI{40}{s} plateau and \SI{10}{s} ramping down). 
The common and independent noise was modelled as coloured Gaussian noise in the \SIrange[range-phrase=\,--\,,range-units = single]{15}{35}{Hz} and \SIrange[range-phrase=\,--\,,range-units = single]{0}{100}{Hz} bands using a second-order zero-phase Butterworth filter. 
To obtain a coefficient of variation (CoV) for the inter-spike intervals of approximately \SI{15}{\%}, unless specified differently, the coefficient of variation (i.e., the standard deviation of the noise divided by the mean drive in the plateau phase) was \SI{20}{\%} for the common noise and \SI{5}{\%} for the independent noise. 
Finally, the input $I_j(t)$ to each motor neuron was a linear combination of the three components.

To quantify the independence of spike trains, we first computed the covariance between all pairs of spike trains.
Next, we computed a spike independence coefficient as the ratio between the sum of the diagonal entries divided by the sum of all elements of the spike train covariance matrix. 
A value of 1 is equivalent to statistically independent spike trains. Lower values indicate dependencies between spike trains.

\subsection{EMG simulation}\label{sec:methods_exp_muaps}
EMG signals were simulated by convolving MUAPs extracted from experimental recordings with the simulated motor neuron spike trains (see Equation~\ref{eqn:convBSSspiketrain}).
Further, we added coloured Gaussian noise with a bandwidth of \SIrange[range-phrase=\,\,to\,\,,range-units = single]{20}{500}{Hz} (using a third-order, zero-phase Butterworth filter) to obtain signal-to-noise ratios (SNR) ranging from \SIrange[range-phrase=\,\,to\,\,,range-units = single]{10}{30}{dB}.
The bandwidth of the noise was chosen to match the bandwidth of hardware filters that are typically implemented in signal amplifiers used for surface EMG recordings.
We extracted a library of MUAPs from the tibialis anterior muscle together with the recruitment thresholds of the motor neurons (in terms of maximum voluntary isometric contractions, MVC) from an existing dataset {\citep{Avrillon2024, Avrillon2023_data}}. 
In that study, 256 electrodes (2x2 64-electrode grids; type: OT Bioelettronica GR04MM1305; inter-electrode distance: \SI{4}{mm}) were used to cover the tibialis anterior muscle, enabling the identification of an extensive number (129 $\pm$ 44 per subject, $n=8$) of spike trains across 10 to \SI{80}{\%} of MVC.
The MUAPs were extracted from multichannel surface EMG signals using spike-triggered averaging with a \SI{\pm 25}{ms} window of each of the provided spike trains. Since the edge samples of the spike-triggered averaged MUAPs resulted in non-zero values, we multiplied them with a Tukey window (cosine fraction 0.1) to ensure compact support. 
Moreover, since the number of MUAPs for the first two subjects (187 and 194) was well above the number of motor neurons used in the motor neuron model, we randomly sampled 300 out of the 381 MUAPs for each simulation.
While the experimental data \citep{Avrillon2024} contains measurements from 4 EMG arrays (64 channels per grid), for the simulations, we only considered 64 channels from one grid (default setting: laterodistal grid).

To study the influence of similar or non-stationary MUAPs systematically, first, we combined the four 64-channel 13x5 grids, resulting in a grid of 26x10 electrodes with \SI{4}{mm} IED. 
Next, linear interpolation was used to obtain a virtual 101x37 grid with \SI{1}{mm} IED.
We obtained two MUAPs of variable similarity by defining a reference MUAP given the original grid indices (i.e., from one 64-channel grid) and a second MUAP by adding an offset value to the reference row or column indices.
Thereby, the MUAP similarity is negatively correlated with the selected row or column offset.
A similar approach was used for obtaining a dictionary of non-stationary MUAPs, i.e., containing the reference MUAP and a variable number of modulated MUAPs with offset values in increments of \SI{1}{mm}. Then each motor neuron spike was associated with one MUAP template of the dictionary (either randomly or as a function of time).

To quantify the difference between two spatio-temporal MUAPs (i.e., considering the data from all channels and each time frame in an integrative sense), we computed for each pair of MUAPs a regularized squared Euclidean (Reg-SqEuc) distance \citep{Farina2008} (in the literature also named energy similarity). 
That is, the mean square difference between two (spatio-temporal) MUAPs regularized by the mean of the energies of the two MUAPs:
\begin{equation}\label{eqn:energy_similarity}
    D_\mathrm{RegSqEuc}^{k,j} \ = \ \frac{\sum\limits_{i}^{M\cdot L} \Big(\hat{\mathbf{h}}_{k}(i) \, - \, \hat{\mathbf{h}}_{j}(i)\Big)^2}{0.5 \Bigg(\sum\limits_{i}^{M \cdot L} \hat{\mathbf{h}}_{k}(i)^2 \, + \,  \sum\limits_{i}^{M \cdot L} \hat{\mathbf{h}}_{j}(i)^2 \Bigg)} \ . 
\end{equation}
Therein, $\hat{\mathbf{h}}_k$ and $\hat{\mathbf{h}}_j$ are the vectorized MUAPs, i.e., obtained by stacking all (spatial and temporal) MUAP components in a 1D vector,  of MU $k$ and MU $j$, respectively. 
Thereby, a value of zero corresponds to identical MUAPs and larger values indicate increasing dissimilarities between MUAPs.
Further, we define the relative MUAP amplitude as the root-mean-square (RMS) of the spatio-temporal MUAPs, divided by the RMS of the total EMG signal.

\subsection{Decomposition based on the MUAP waveforms}
After extending the simulated signals (default: $R=16$), we used ZCA whitening to decorrelate the extended data.
To limit the effect of noise, we added a regularization factor to the eigenvalues that is equal to the mean of the lowest half of the eigenvalues. 
In Section~\ref{sec:identifiability}, we have shown that for robust learning, the projection vector converges to a scaled version of the extended and whitened MUAP.
Thus, to avoid any bias towards a specific decomposition algorithm, we directly reconstructed the (delayed) spike trains of MU $k$ given the known MUAP waveforms:
\begin{equation}
\label{eqn:optimalfilter}
\begin{split}
 \mathbf{\widehat{\tilde{s}}}_{k}(t,l) \ = \ \frac{\mathbf{\tilde{h}}_{kl}^T}{|\mathbf{\tilde{h}}_{kl}|} \, \mathbf{\tilde{z}}(t) \ , \ \text{with} \ \  0 < l < L+R-1 \ .
\end{split}    
\end{equation}
Therein $\mathbf{\widehat{\tilde{s}}}_{k}(t,l)$ are delayed estimates of the $k$th spike train, $\mathbf{\tilde{z}}(t)$ is the whitened extended observation matrix, and $\mathbf{\tilde{h}}_{kl}$ are the columns of the extended and whitened mixing matrix (see Equation~\ref{eqn:MUAP_filter}). 
Note that $\mathbf{\widehat{\tilde{s}}}_{k}(t,l)$ is different from Equation~\eqref{eqn:convBSSwk} since it includes $(L+R-1)$ delayed sources where the source with the highest skewness was selected as the representative source of all the delayed spike trains.
\subsection{Uncertainty of reconstructed spike trains}\label{sec:seperability_metric}

\begin{figure}[ht!]
\centering
\includegraphics[width=\textwidth]{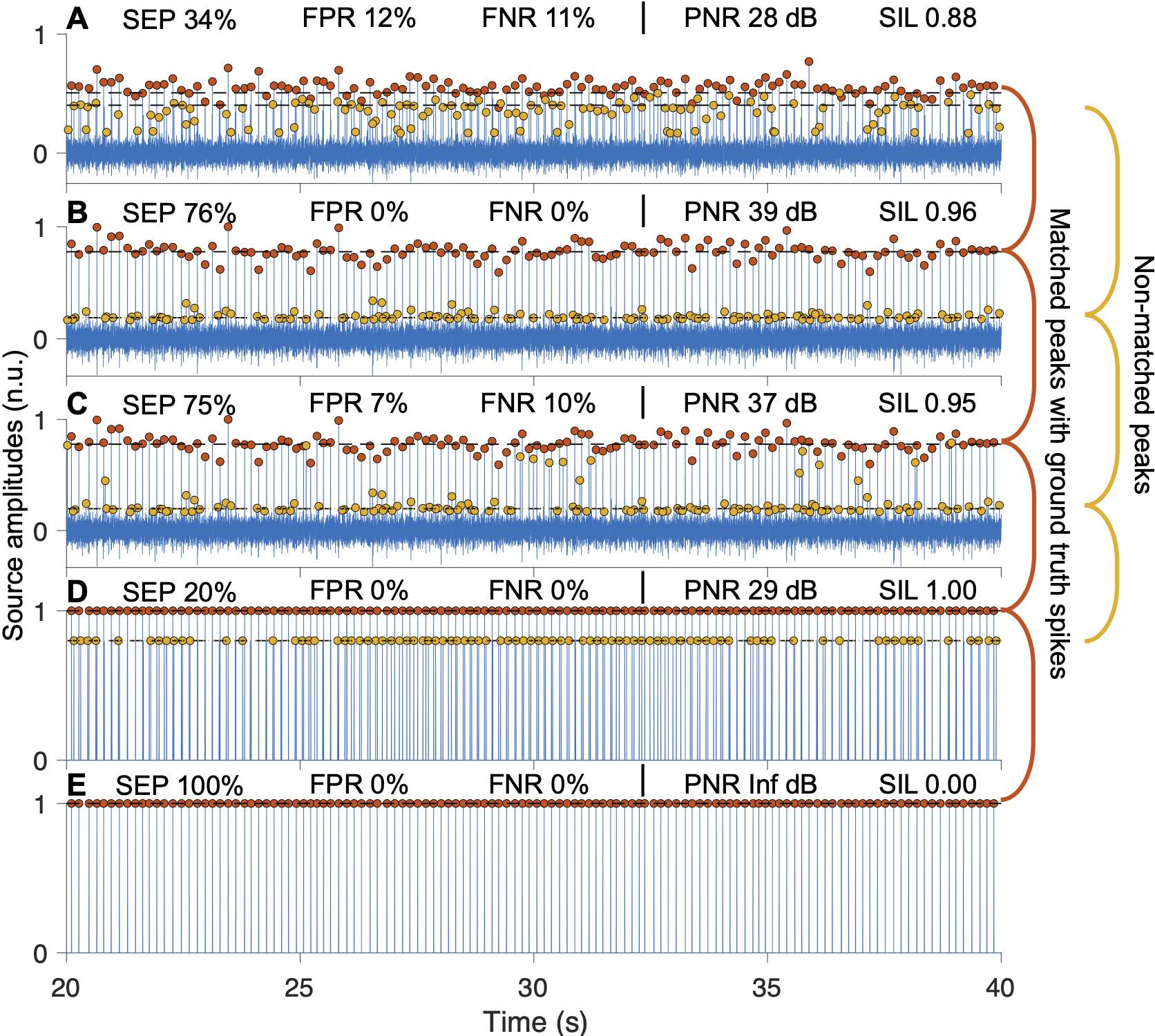}
\caption{
Schematic illustration of the behavior of different source quality metrics regarding the identifiability of motor unit (MU) sources. Panels \textbf{A} to \textbf{E} represent different sources with different degrees of separability (SEP), false positive rates (FPR), false negative rates (FNR), pulse-to-noise ratios (PNR), and silhouette (SIL) values. Note that sources with no baseline noise (\textbf{D}-\textbf{E}) are not realistic experimentally, but these limiting cases provide insights into the pitfalls of different metrics.
}
\label{fig:Separabilitymetric}
\end{figure}

To evaluate the identifiability of spike trains, we empirically computed the relative peak separation (see Equation~\ref{eqn:separation_of_MUs}).
To do so, $\widehat{\tilde{s}}_k(t)$ was normalized to a maximal value of 1 and positive skewness. 
Next, a peak detection method identified all positive peaks above three times the mean absolute deviation.
The lag between the ground truth spike train and the estimated spike train was found using cross-correlation. 
After alignment, we clustered the peaks into matched (i.e., true spikes) and non-matched (i.e., false spikes) with the ground truth spikes (within a $\pm 5$ ms window due to estimated spike train variability). 
We truncated the number of background peaks to ensure that the number of peaks in the matched and non-matched clusters was identical. 
Finally, we computed the relative peak separation as the difference between the median of the matched and non-matched peak amplitudes divided by the median of the matched peak amplitudes {(Figure~\ref{fig:Separabilitymetric}A-E)}. 
Since this metric can result in negative values (false spikes larger than true spikes), we truncated it such that the minimum value is zero. 

The relative peak separation does not consider noise. 
Thus, we added an estimate of the false positive and false negative rates. The false positive rate was defined as the number of unmatched peaks above the 5th percentile of the matched peaks. The false negative rate was defined as the number of matched peaks below the 95th percentile of the unmatched peaks. The separability, false positive, and false negative metrics, together with PNR and SIL values, are exemplified for different sources {(Figure~\ref{fig:Separabilitymetric}A-E)}.
Note that while the separability, false positive rate and false negative rate are based on the ground truth spikes, PNR and SIL blindly cluster the spikes.  
It can be seen that PNR and SIL have limited sensitivity to the false positive and false negative rates {(Figure~\ref{fig:Separabilitymetric}B-C)}, as shown in Section~\ref{sec:source_quality}.
Further, the artificial example shown in Figure~\ref{fig:Separabilitymetric}D illustrates that SIL is not directly correlated with the separation between true spikes and background spikes.
Note that the SIL becomes zero for a perfect spike train because the true peaks are divided into two clusters with identical mean value {(Figure~\ref{fig:Separabilitymetric}E)}. 
Assigning an amplitude of zero to the background peak cluster, one would obtain SIL equal to one.

\subsection{Data processing and statistics}
The conducted in silico experiments are three-fold: (i) exemplary simulations; (ii)  conditional simulations, where a maximum number of two parameters (e.g., MUAP amplitude and MUAP similarity) is systematically varied, showcasing the specific influence of the modulated parameters on various metrics, e.g., separability, false positive rate, false negative rate, regularized squared Euclidean (Reg-SqEuc) distance between MUAPs, PNR, SIL, total number of decomposed MUs etc; (iii) series of random model configurations to show correlation between multiple factors. 
Each conditional simulation was repeated multiple times, and we report either binned fractions of decomposed MUs or empirical percentiles of different performance metrics.
Further, we used Pearson's linear correlation coefficient to quantify the correlation between different parameters (see Section~\ref{sec:quality_source_metric} and \ref{sec:results_population}). The simulations, decompositions, and data processing were performed using MATLAB, where the source code to regenerate the results and figures is available in a git repository \citep{KlotzRohlen2025}.

\section{Results}

\subsection{Simulated vs experimental EMG signals}\label{sec:results_accuracy}
Computer simulations enable unique insights into the output of MU decompositions that are hardly feasible experimentally. 
To demonstrate the plausibility and realism of the simulated signals, we first conducted a comparison of simulated signals and experimental data.
Therefore, we simulated a 256-channel surface EMG signal (four 64-channel grids like the experimental data used in this study) using a trapezoid-shaped drive with a maximal amplitude of \SI{13}{nA} (\SI{10}{s} ramping up, \SI{40}{s} plateau and \SI{10}{s} ramping down).
The common noise had a CoV of \SI{10}{\%} and the independent noise a CoV of \SI{10}{\%}. This resulted in an active pool of 163 MUs. Finally, we added coloured Gaussian noise to the EMG signal, resulting in a SNR of \SI{20}{dB}.

\begin{figure}[ht]
        \centering
        \includegraphics[width=1.0\linewidth]{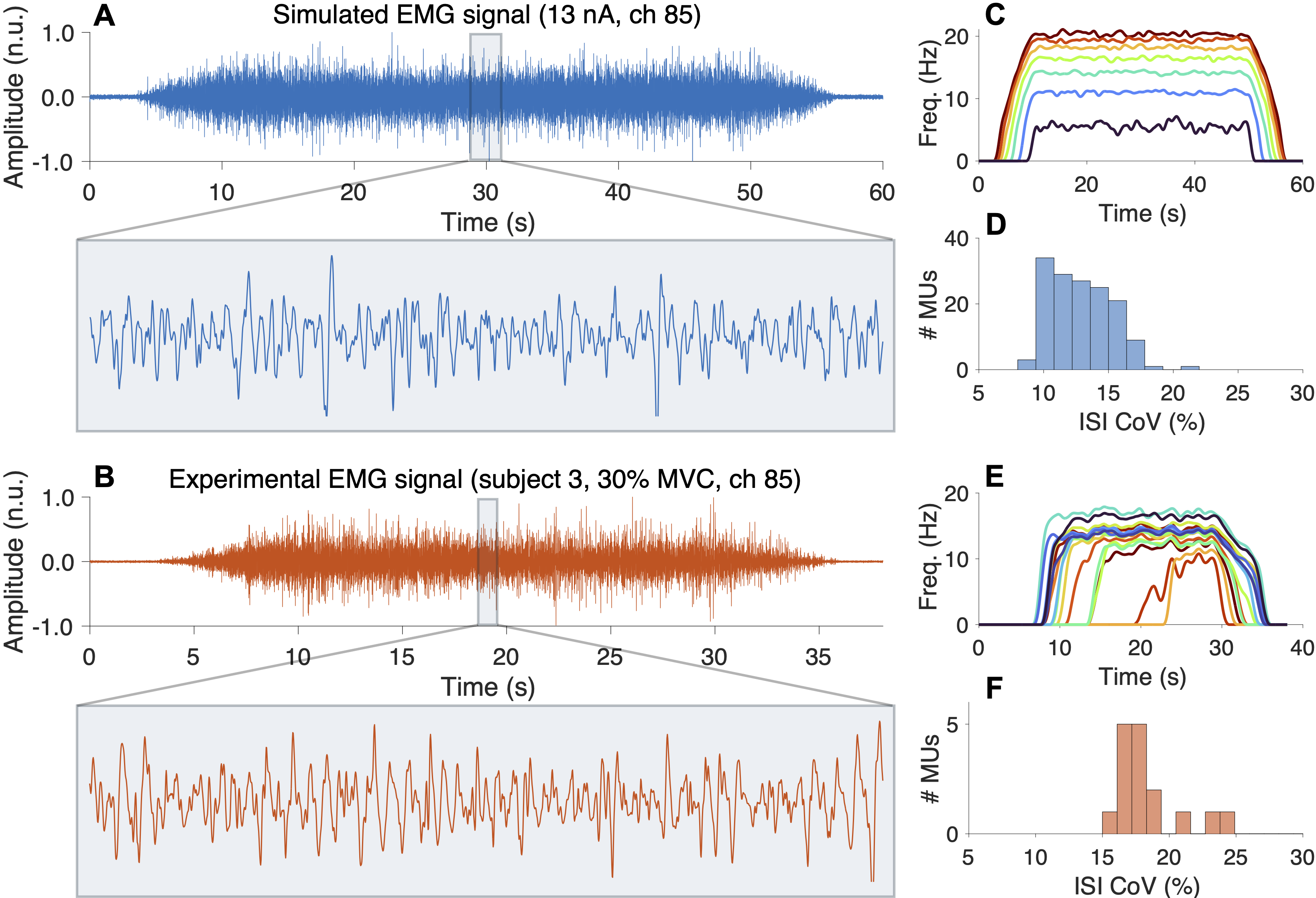}
        \caption{Simulated vs experimental EMG signals. (\textbf{A}) An example of a simulated EMG signal (channel 85) for a trapezoid contraction with a mean drive of \SI{13}{nA}, \SI{20}{dB} coloured Gaussian noise, CCoV and ICoV equal to \SI{10}{\%} (n=163 active MUs). (\textbf{B}) An example of an experimental EMG signal (channel 85) for subject 3 at \SI{30}{\%} MVC. (\textbf{C}) Simulated data: The instantaneous frequencies of MU numbers 1, 25, 50, 75, 100, 125, and 150. (\textbf{D}) Simulated data: The distribution of inter-spike-interval coefficient of variations (ISI CoV) for the MU pool. (\textbf{E}) Experimental data: The instantaneous frequencies of 17 detected MUs. (\textbf{F}) Experimental data: The distribution of ISI CoV for the identified MUs.}
        \label{fig:emg_signal_examples}
    \end{figure}

An exemplary simulated EMG signal for channel 85 {(Figure~\ref{fig:emg_signal_examples}A)} is visually similar to an exemplary experimental EMG signal (subject 3 at \SI{30}{\%} MVC) for channel 85 {(Figure~\ref{fig:emg_signal_examples}B)}. Note the different signal and ramp lengths. The median frequency for these signals, based on 10 seconds in the middle of the contraction during the steady part, was 54.9 Hz and 54.1 Hz, respectively. Considering all 256 channels, the median frequency of the simulated signals was 51 $\pm$ 4.4 Hz, whereas it was 56.6 $\pm$ 6.1 Hz for the experimental signals. The instantaneous firing rates for the simulated signals (exemplified through MU numbers 1, 25, 50, 75, 100, 125, and 150), ranged from \SI{20}{Hz} down to \SI{5}{Hz} {(Figure~\ref{fig:emg_signal_examples}C)} were within the range of the $n=17$ MUs identified in the experimental signals {(Figure~\ref{fig:emg_signal_examples}E)}. The distribution of ISI CoV for the MU pool was also within a reasonable range, around \SIrange[range-phrase=\,\,to\,\,,range-units = single]{10}{20}{\%} {(Figure~\ref{fig:emg_signal_examples}D)}, where it was \SIrange[range-phrase=\,\,to\,\,,range-units = single]{15}{25}{\%} for the experimentally identified MUs {(Figure~\ref{fig:emg_signal_examples}F)}.

\subsection{Identification errors}\label{sec:results_accuracy}
\begin{figure}[ht!]
\centering
\includegraphics[width=\textwidth]{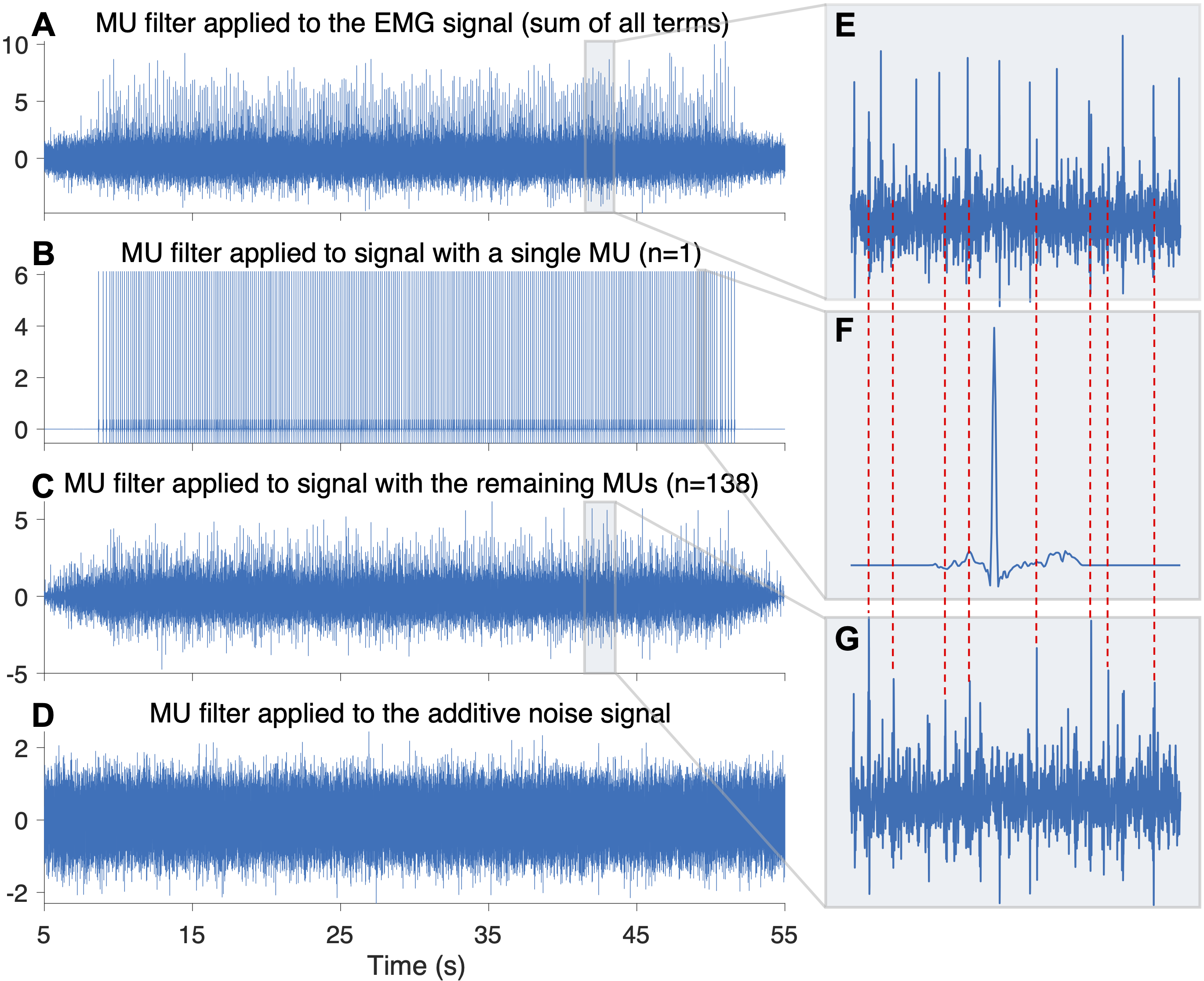}
\caption{Exemplary illustration of the reconstruction errors of one motor unit spike train (MU \#105) from 64-channel surface electromyography (EMG) where 139 MUs were active. The projection vector (calculated based on the original MUAP waveforms) was applied to the simulated EMG signals (\textbf{A} and \textbf{E}), the signal contributions of MU \#105 (\textbf{B} and \textbf{F}), the remaining 138 MUs (\textbf{C} and \textbf{G}), and to the additive coloured noise (\textbf{D}). Note that the reconstructed MU source from the full EMG signal (\textbf{A}) is equal to the superposition of the isolated projected signal components \textbf{B}-\textbf{D}.
}
\label{fig:reconstruction_accuracy}
\end{figure}

To investigate how a reconstructed motor neuron spike train is affected by different contributions of the total (extended and whitened) EMG signal, we simulated an exemplary 64-channel surface EMG signal. 
We used a trapezoid-shaped drive with a maximal amplitude of \SI{10}{nA} (\SI{10}{s} ramping up, \SI{40}{s} plateau and \SI{10}{s} ramping down).
The common noise had a CoV of \SI{30}{\%} and the independent noise a CoV of \SI{3}{\%}.
This resulted in an active pool of 139 MUs (mean firing rate $10.7 \pm 4.3\,$Hz).
Finally, we added coloured Gaussian noise to the EMG signal such that we obtained a SNR of \SI{20}{dB}. 
After selecting an exemplary MU (i.e., \#105), we applied the optimal projection vector (i.e., the extended and whitened MUAP waveforms) to the simulated EMG signals to obtain a MU source (Figure~\ref{fig:reconstruction_accuracy}A). 
For the simulated data, the estimated source can be split up into the spiking activity of interest and noise contributions. We illustrated these noise sources by applying the MU filter to the isolated contribution of MU \#105 {(Figure~\ref{fig:reconstruction_accuracy}B)}, the contribution of the remaining 138 MUs {(Figure~\ref{fig:reconstruction_accuracy}C)}, and the additive coloured noise {(Figure~\ref{fig:reconstruction_accuracy}D)}.
It can be observed that the projection of the single MU signal closely approximates a Dirac delta function (Figure~\ref{fig:reconstruction_accuracy}B).
The spike amplitude was $6.12$, which was equivalent to the norm of the corresponding column in the extended and whitened mixing matrix (i.e., as predicted by Equation~\ref{eqn:source_amplitude_estimate}).
The approximation error is because the delayed versions of the extended and whitened MUAP are not fully independent (Figure~\ref{fig:reconstruction_accuracy}A). 
Considering a window of plus/minus one sample, the observed spike amplitude was, on average, \SI{68}{\%} of the spike's peak height.
The projection error was below \SI{15}{\%} for all other delays.
Notably, this allows the identification of doublets, i.e., two consecutive discharges that are about \SIrange[range-phrase=\,\,to\,\, ,range-units = single]{2}{10}{ms} apart {\citep{Christie2006}}, if two consecutive spikes are separated by at least three samples; see {Supplementary Figure~C2}.
The amplitude of the largest peak due to the activity of other MUs was $6.15$ (Figure~\ref{fig:reconstruction_accuracy}C and G), considerably larger than the expected amplitude of the most dominant background source (amplitude: 3.63).
Further, the variance of the projected noise  (Figure~\ref{fig:reconstruction_accuracy}D) was $0.30$, i.e., considerably lower than the worst-case estimate, which is a variance equal to one.
Nevertheless, although the simulated noise is Gaussian and we truncated the eigenvalues for constructing the whitening matrix, the influence of the projected noise can not be fully eliminated.
Note that the reconstructed MU source from the full EMG signal {(Figure~\ref{fig:reconstruction_accuracy}A)} is equal to the superposition of the isolated projected signal components {(Figure~\ref{fig:reconstruction_accuracy}B-D)}, thus verifying Equation~\eqref{eqn:convBSSallerrors}.

\subsection{Similiar MUAPs}\label{sec:results_similar_muaps}
A core assumption in the separability of MUs in surface EMG signals is the MUAP uniqueness. 
The theory presented in Section~\ref{sec:existence} demonstrates that MU identifiability also depends on the whitening, the MUAP amplitude, and noise. 
To test this theoretical prediction, we conducted a series of in silico experiments. 
We used a trapezoidal-shaped drive (ramp up: \SI{10}{s}; plateau: \SI{40}{s}; ramp down: \SI{10}{s}; max. drive: \SI{7}{nA}; CoV of the common noise: \SI{20}{\%}; CoV of the independent noise: \SI{5}{\%}), resulting in a pool of 75 MUs, where we considered two MUs (\#50 and \#49), for which we modulated the MUAP amplitude and similarity (see Section~\ref{sec:methods_exp_muaps}). 
To study the effect of different noise levels, we considered SNR values of \SI{20}{dB} and \SI{10}{dB}, respectively.

\begin{figure}[ht!]
\centering
\includegraphics[width=\textwidth]{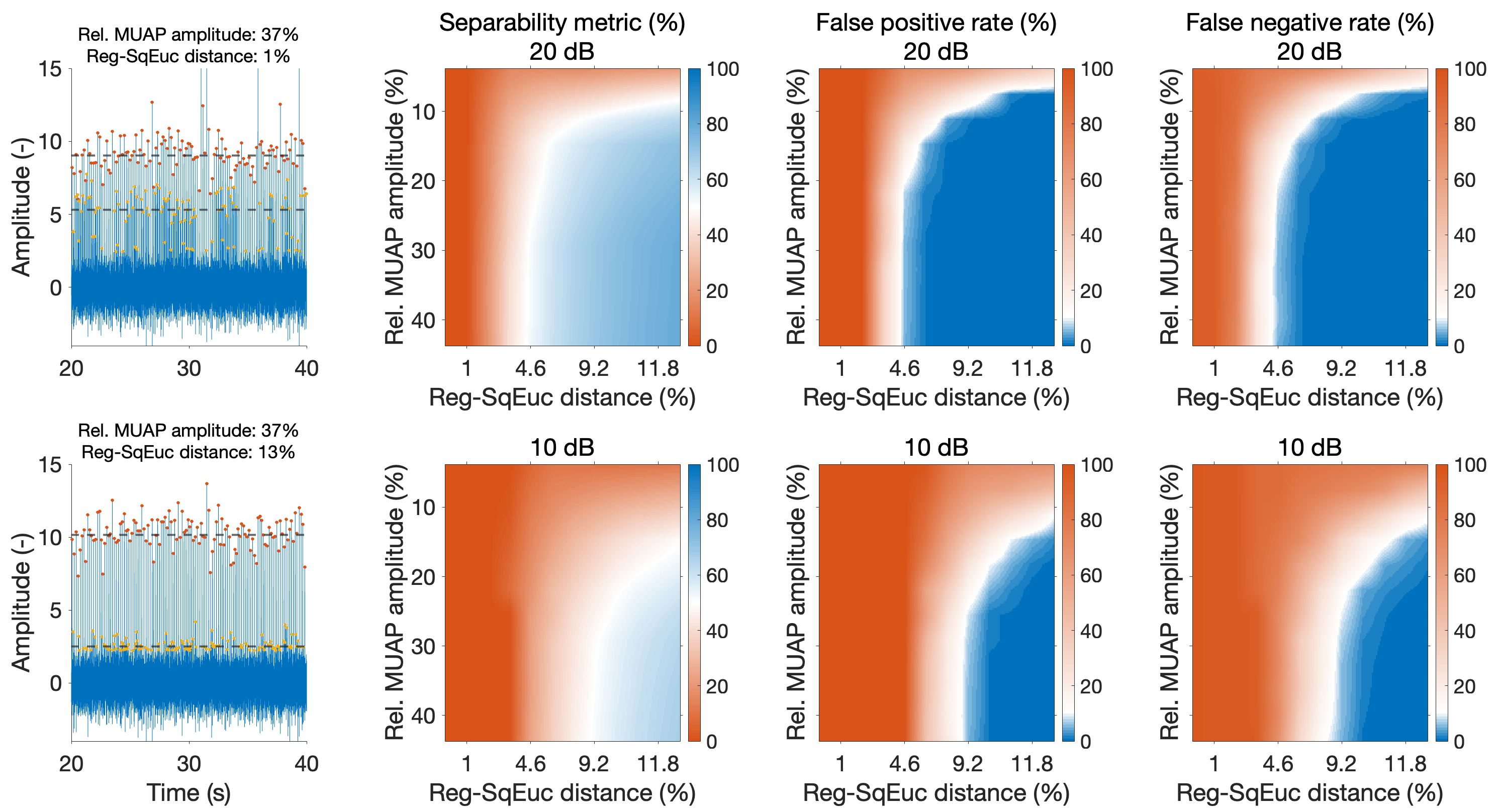}
\caption{Separability between two MUs (\#49 and \#50) depending on MUAP similarity and amplitude.
The left column shows two exemplarily selected estimated spike trains (red: matched spikes, yellow: unmatched spikes, dotted lines: median values of matched and unmatched spikes).
Further, columns two to four show for two different signal-to-noise ratio (SNR) values (\SI{20}{dB} and \SI{10}{dB}) the separability metric, the false-positive rate, the false-negative rate depending on the regularized squared Euclidean (Reg-SqEuc) distance between the two MUAPs (x-axis) and the relative MUAP amplitudes (y-axis). The relative MUAP amplitude is defined as the root-mean-square (RMS) of the spatio-temporal MUAPs divided by the RMS of the total EMG signal. Note that the colormaps for the false positive and false negative rates were constructed such that the white zone is about \SI{10}{\%}.
}
\label{fig:similar_muaps}
\end{figure}

The estimated spike trains shown in Figure~\ref{fig:similar_muaps} (left column) exemplarily illustrate that the MU spikes become indistinguishable when two MUAPs have increased similarity. Both sources have the same relative MUAP amplitude (the RMS of the spatio-temporal MUAPs divided by the RMS of the total EMG signal).
In detail, it is observed that the identifiability of MUs depends on the MUAP similarity and amplitude as well as noise level (Figure~\ref{fig:similar_muaps}, columns two to four). 
For the case with an SNR of \SI{20}{dB}, the minimal Reg-SqEuc distance that allows a robust decomposition, i.e., where both false positive and negative rates are below \SI{10}{\%}, is \SI{4.6}{\%}. Further, the MUAP amplitude can be decreased to \SI{60}{\%} until the false positive rate and false negative rates quickly increase.
Further, it is observed that increasing the noise level shifts the identifiability ranges towards higher Reg-SqEuc distance values and amplitudes. 
For an SNR of \SI{10}{dB}, the MU spike train could be robustly identified for Reg-SqEuc distance values larger than \SI{9.2}{\%}. This value quickly increases if the scaled amplitude drops below \SI{70}{\%}.

To test the robustness of these observations, we repeated this in silico experiment for 21 model configurations (SNR values ranged from 15 to \SI{30}{dB}, and the mean drive ranged from 7 to \SI{9}{nA}).
Considering all MUAPs with minimal Reg-SqEuc distance between 10 and \SI{15}{\%} and relative amplitudes between 20 and \SI{25}{\%}, the fraction of identifiable MUs is \SI{97.2}{\%} (Table~\ref{tab:results_similar_muaps}).
The fraction of detectable MUs decreased to \SI{68.4}{\%} when the Reg-SqEuc distance was between 2.5 and \SI{5}{\%} (relative MUAP amplitudes between 20 and \SI{25}{\%}). The detectable MUs decreased further to only \SI{23.5}{\%} for the same Reg-SqEuc distance but with a smaller relative MUAP amplitude (10 to \SI{15}{\%}).

\begin{table}[hb!]
\footnotesize
 \centering
 \caption{
 Fraction of detectable sources (from 21 in silico experiments) in discrete bins depending on the regularized squared Euclidean (Reg-SqEuc) distance to the most similar MUAP (columns) and the relative MUAP amplitude (rows).
 }
 \label{tab:results_similar_muaps}
 \begin{tabular}{cccccc}
        \toprule
        & \multicolumn{5}{c}{\textbf{Minimum Reg-SqEuc distance}} \\
        \cmidrule(lr){2-6}
        \textbf{Rel. MUAP Amplitude} & 0 to \SI{2.5}{\%} & 2.5 to \SI{5}{\%} & 5 to \SI{7.5}{\%} & 7.5 to \SI{10}{\%} & 10 to \SI{15}{\%}\\
 \midrule
  0 to \SI{10}{\%} & 0.0 & 0.0 & 5.5 & 8.9 & 15.6\\[1mm]
  10 to \SI{15}{\%} & 0.0 & 23.5 & 55.6 & 55.2 & 80.0\\[1mm]
  15 to \SI{20}{\%} & 16.7 & 35.5 & 67.7 & 77.4 & 86.7\\[1mm]
  20 to \SI{25}{\%} & 42.9 & 68.4 & 87.1 & 96.2 & 97.2\\
 \bottomrule
 \end{tabular}
 \end{table}

\subsection{Spike train dependencies}\label{sec:results_independence}

To investigate the separability of MUs when varying the degree of spike train dependence, we performed a series of simulations using a trapezoid-shaped drive (ramp up: \SI{10}{s}; plateau: \SI{40}{s}; ramp down: \SI{10}{s}; max. drive: \SI{7}{nA}) and an SNR of \SI{15}{dB}.
However, in contrast to the previous examples, the CoV of the common noise was varied from \SIrange[range-phrase=\,\,to\,\,,range-units = single]{0}{60}{\%}, and the CoV of the independent noise was varied from \SIrange[range-phrase=\,\,to\,\,,range-units = single]{0}{20}{\%}.
Note that increasing the CoV of the common noise and decreasing the CoV of the independent noise increases the dependency between the spike trains (although the spike trains are never identical due to the difference in motor neuron properties).
Given the covariance matrix of the spike trains, for a CoV of \SI{0}{\%} of both common noise and independent noise, the spike independence coefficient was \SI{94}{\%}.
Increasing the common noise to \SI{60}{\%}, the spike independence coefficient decreased to \SI{43}{\%} (CoV of the independent noise: \SI{0}{\%}) and \SI{51}{\%} (CoV of the independent noise: \SI{20}{\%}).
We further note that increasing the input noise increased the size of the active MU pool.
To isolate the effect of spike train correlations, we truncated the size of the active MU pool to match the value given a CoV of \SI{0}{\%} (both for the independent and common noise), i.e., 58 MUs.

\begin{figure}[hb!]
\centering
\includegraphics[width=\textwidth]{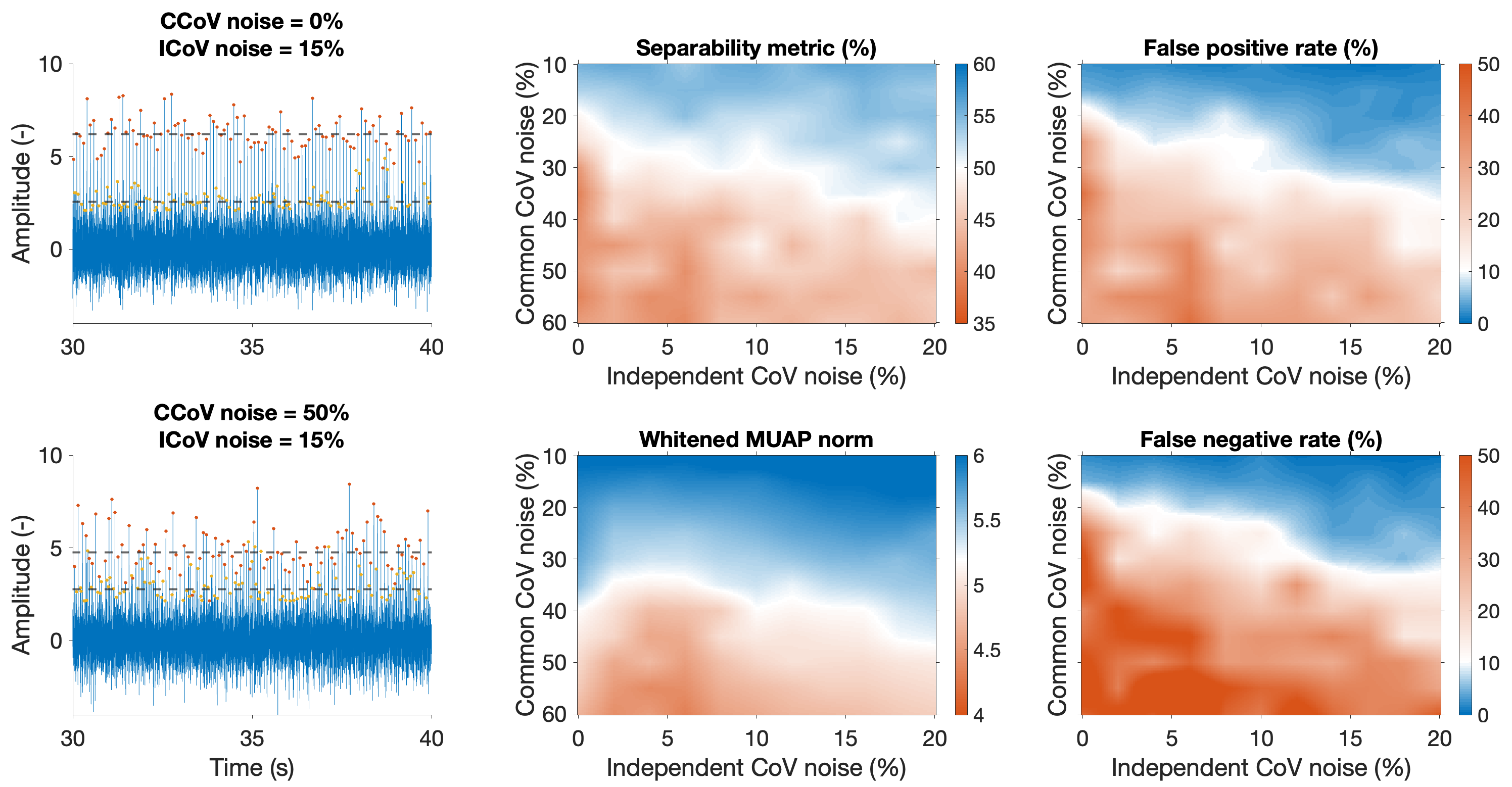}
\caption{Influence of spike train correlations on motor unit (MU) identifiably for one exemplary selected MU (\#1). This is achieved by changing the coefficient of variation (CoV) of the common noise (CCoV) and independent noise (ICoV) of the motor neuron pool input current. Left column: Two exemplary selected estimated spike trains; second column: Separability metric and norm of the whitened MUAP depending on the CCoV and ICoV;
right column: False-positive rate and false negative rate depending on the CCoV and ICoV.
Note that the colormaps for the false positive and false negative rates were constructed such that the white zone is about \SI{10}{\%}.}
\label{fig:common_spikes}
\end{figure}

Figure~\ref{fig:common_spikes} exemplarily showcases the decomposition performance of one MU (\#1). 
We found that separability was the lowest, and false positive and negative rates were the highest when the MUs were the most synchronised. 
Yet, there was a clear tendency that the CoV of common noise had a larger effect than the CoV of independent noise.
Notably, it is observed that the separability metric and the norm of the whitened MUAPs are correlated with the CoV of the common and independent noise.
Both the separability metric and the whitened MUAP's norm were the lowest when the spike trains had the lowest spike independence coefficient.
Moreover, there was a clear tendency for the common CoV noise to have a larger effect on the norm changes than the independent CoV noise.

To investigate the robustness of the observed behavior, the same in silico experiments was repeated for 21 model configurations (SNR values ranged from 15 to \SI{30}{dB}, and the mean drive ranged from 7 to \SI{9}{nA}), where the active MU pool size ranged from 58 MUs to 125 MUs.
The highest number of MU spike trains (median value: 15.0) was reconstructed when considering the maximum value (\SI{20}{\%}) of the ICoV noise and the minimal value (\SI{10}{\%}) of the CCoV (Table~\ref{tab:results_spike_train_corr}). 
When reducing the ICoV to \SI{0}{\%} and increasing the CCoV to \SI{50}{\%}, the median number of identified MU spike trains was 1.

\begin{table}[t]
\footnotesize
 \centering
 \caption{
 Number of MUs (10/50/90 percentile) with a false-positive rate below \SI{10}{\%} for 21 virtual subjects depending on the coefficient of variance (CoV) of the independent (ICoV) and common (CCoV) noise.
 }
 \label{tab:results_spike_train_corr}
 \begin{tabular}{ccccccc}
        \toprule
        & \multicolumn{6}{c}{\textbf{ICoV}} \\
        \cmidrule(lr){2-7}
        \textbf{CCoV} & \textbf{\SI{0}{\%}} & \textbf{\SI{4}{\%}} & \textbf{\SI{8}{\%}} & \textbf{\SI{12}{\%}} & \textbf{\SI{16}{\%}} & \textbf{\SI{20}{\%}} \\
 \midrule
  \textbf{\SI{10}{\%}} & 4.2/13.0/31.0 & 3.8/15.0/29.8 & 4.2/15.0/31.0 & 3.8/15.0/31.2 & 3.8/16.0/31.8 & 3.8/15.0/32.0\\[1mm]
  \textbf{\SI{30}{\%}} & 0.0/6.0/13.4 & 0.6/6.0/17.4 & 0.6/7.0/19.8 & 1.0/7.0/19.8 & 1.0/8.0/22.0 & 1.0/10.0/22.8\\[1mm]
  \textbf{\SI{50}{\%}} & 0.0/1.0/6.0 & 0.0/3.0/8.0 & 0.0/3.0/9.4 & 0.0/4.0/12.2 & 0.0/4.0/13.0 & 0.0/4.0/14.2\\[1mm]

 \bottomrule
 \end{tabular}
 \end{table}

We also investigated the degree of separability when two MUs had identical spike trains but where one of the spike trains was temporarily shifted (\SIrange[range-phrase=\,\,to\,\,,,range-units = single]{1}{10}{ms}).
This artificially induces a strong dependence between the two sources since knowing one spike train can fully explain the other one. 
We found that for some MUs, it did not affect their identifiability as long as they had a higher norm {(Supplementary Figure~C3)}. 
On the other hand, the MU with a lower MUAP norm was separable with minimal to no false positive and negative rates for delays larger than \SI{5}{ms}. 
For shorter delays, the false positive and negative rates increased rapidly.

\subsection{Non-stationary signals}
Signal stationarity is a critical assumption for convolutive BSS.
It has been observed \citep[e.g.,][]{Rohlen2025} that classic convolutive BSS is capable of handling modest non-stationarities by learning a projection vector representing an average MUAP (see Supplementary Figure C4).
Here, we systematically varied the degree of non-linearity by first constructing a dictionary of non-stationary MUAPs (see Section~\ref{sec:methods_exp_muaps}) and then modulating the range of considered MUAP templates (from 1, i.e., the stationary case, to 19 modulated MUAPs) given the same input spike train.
Thereby, each spike was randomly assigned one of the considered spatio-temporal MUAPs.

\begin{figure}[ht!]
\centering
\includegraphics[width=1.0\textwidth]{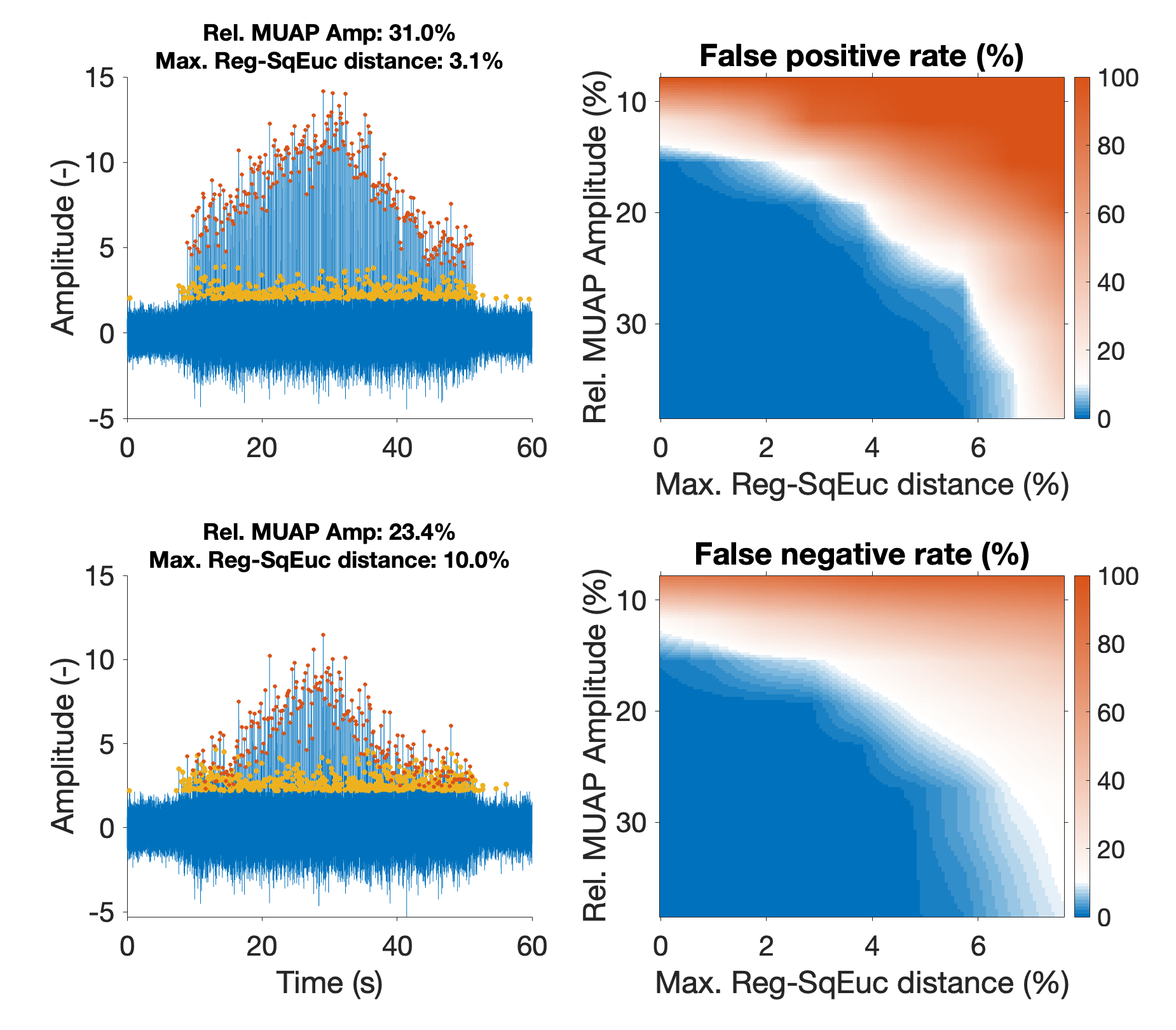}
\caption{Investigation of motor unit (MU) identifiability considering MUs associated with non-stationary MUAPs for an exemplarily selected MU (\# 50). Left column. Two exemplary selected estimated MU spike trains with different MUAP amplitudes and a different level of signal non-stationarity. The temporal evolution of the MUAPs is linear.  Right column: False-positive rate and false-negative rate depending on the regularized squared Euclidean (Reg-SqEuc) distance to the most dissimilar modulated MUAP waveform and the relative MUAP amplitude.}
\label{fig:non-stationary}
\end{figure}

To illustrate how non-stationarities, i.e., varying MUAP shapes between the discharges of the same motor neuron, affect the motor neuron identification, we simulated a trapezoidal-shaped contraction (ramp up: \SI{10}{s}; plateau: \SI{40}{s}; ramp down: \SI{10}{s}; max. drive: \SI{7}{nA}; CoV of the common noise: \SI{20}{\%}; CoV of the independent noise: \SI{5}{\%}) and assumed an SNR of \SI{20}{dB}.
Figure~\ref{fig:non-stationary} shows that, depending on the degree of non-linearity, the MU spikes are still detectable. 
However, with increasing MUAP variability and decreasing MUAP amplitude, the false positive rate and the false negative rate increase.
Thereby, the effect of false-positive spikes dominates that of false-negative spikes.
For example, for a maximal value of the Reg-SqEuc distance between the median MUAP and the most different MUAP of \SI{3.1}{\%} and a relative MUAP amplitude of \SI{31.0}{\%}, the false-positive rate is \SI{0.0}{\%} and the false-negative rate is \SI{0.4}{\%}.
Given a maximal value of the Reg-SqEuc distance of \SI{7.6}{\%} and a relative MUAP amplitude of \SI{23.4}{\%}, the false positive rate is \SI{67.9}{\%} and the false negative rate is \SI{17.7}{\%}.

We repeated this in silico experiment for 21 model configurations (Table~\ref{tab:non-stationary}).
Thereby, the mean cortical drive varied between \SIrange[range-phrase=\,\,and\,\,,range-units = single]{7}{9}{nA} and the noise was varied between \SIrange[range-phrase=\,\,and\,\,,range-units = single]{15}{30}{dB}.
Further, the maximum Reg-SqEuc distance between the reference MUAP and the most dissimilar modulated MUAP varied between \SIrange[range-phrase=\,\,and\,\,,range-units = single]{4.8}{48.4}{\%}.
Given a maximal value of the Reg-SqEuc distance between the reference MUAP and the most dissimilar MUAP waveform between \SIrange[range-phrase=\,\,and\,\,,range-units = single]{0}{2.5}{\%} and a relative MUAP amplitude between \SIrange[range-phrase=\,\,and\,\,,range-units = single]{20}{25}{\%}, the binned fraction of decomposed MUs was \SI{91.1}{\%}.
Considering lower MUAP amplitudes or conditions with higher MUAP variability, the fraction of detectable MUs decreases. 
Interestingly, the influence of MUAP non-stationarities was larger for high-amplitude MUAPs than for low-amplitude MUAPs. 
For example, considering MUs with relative MUAP amplitudes between \SIrange[range-phrase=\,\,and\,\,,range-units = single]{20}{25}{\%} and comparing bins with a maximal Reg-SqEuc distance ranging from \SIrange[range-phrase=\,\,to\,\,,range-units = single]{0}{2.5}{\%} and \SIrange[range-phrase=\,\,to\,\,,range-units = single]{10}{15}{\%}, the fraction of detectable MUs decreased by a factor of 3.04.
Given relative MUAP amplitudes between \SIrange[range-phrase=\,\,and\,\,,range-units = single]{10}{15}{\%}, the fraction of detectable MUs decreased only by a factor of 1.90.

\begin{table}[hb!]
\footnotesize
 \centering
 \caption{
 Fraction of detectable sources from non-stationary signals (pooled from 21 in silico experiments) in discrete bins depending on the regularized squared Euclidean (Reg-SqEuc) distance to the most disimilar MUAP (columns) and the relative MUAP amplitude (rows).
 }
 \label{tab:non-stationary}
 \begin{tabular}{cccccc}
        \toprule
        & \multicolumn{5}{c}{\textbf{Maximum Reg-SqEuc distance}} \\
        \cmidrule(lr){2-6}
        \textbf{Rel. MUAP Amplitude} & 0 to \SI{2.5}{\%} & 2.5 to \SI{5}{\%} & 5 to \SI{10}{\%} & 10 to \SI{15}{\%} & 15 to \SI{20}{\%}\\
 \midrule
  0 to \SI{10}{\%} & 30.9 & 24.1 & 21.3 & 14.0 & 6.8\\[1mm]
  10 to \SI{15}{\%} & 77.9 & 67.1 & 64 & 41.1 & 42.9\\[1mm]
  15 to \SI{20}{\%} & 80.5 & 62.3 & 69.6 & 38.1 & 27.6\\[1mm]
  20 to \SI{25}{\%} & 91.1 & 53.5 & 50.0 & 30.0 & 0.0\\
 \bottomrule
 \end{tabular}
 \end{table}

\subsection{Extension factor}\label{sec:results_extension}
\begin{figure}[ht!]
\centering
\includegraphics[width=\textwidth]{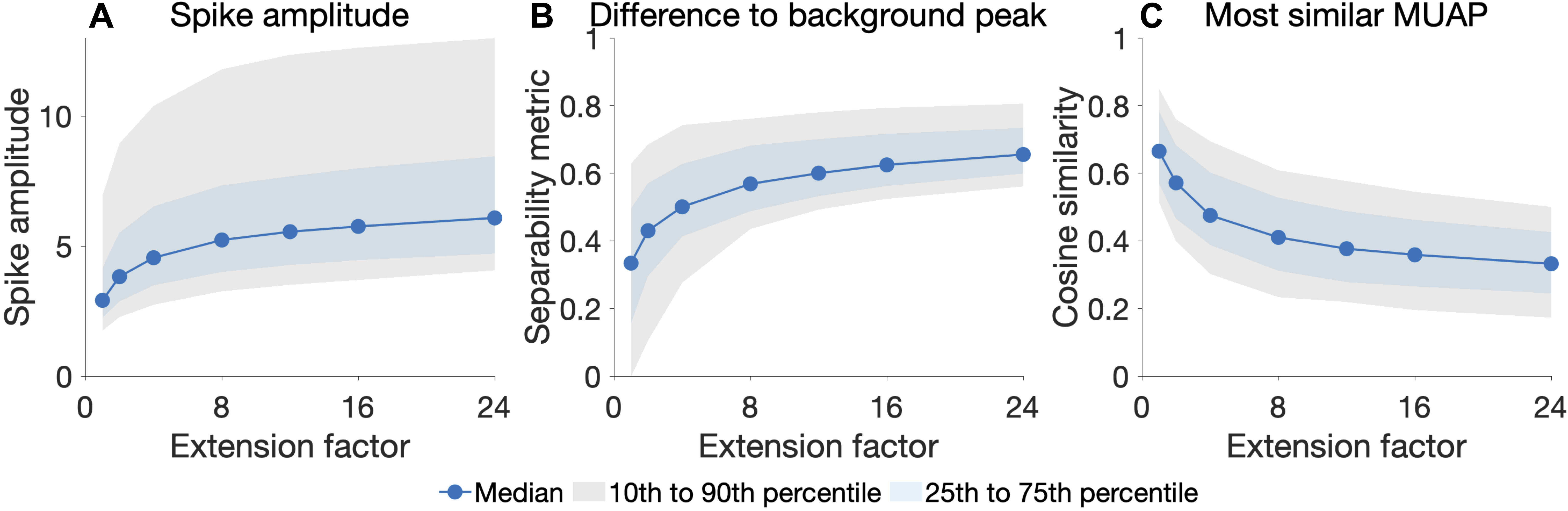}
\caption{Influence of the extension factor $R$ on the decomposition of a population of 372 motor units (MUs). 
Effect on the expected spike amplitude (\textbf{A}). Influence on the relative peak separation relative to the largest background peak (\textbf{B}). Effect on the cosine similarity of the most similar extended and whitened MU action potential (MUAP) (\textbf{C}).}
\label{fig:extension_factor}
\end{figure}

To test the importance of extension factor $R$, we investigated for a population of 372 motor units from 20 virtual recordings how changing the extension factor affected the spike amplitudes, relative peak separation, and the cosine similarity of the most similar MUAP (Figure~\ref{fig:extension_factor}A-C). By using a trapezoid-shaped drive (ramp up: \SI{10}{s}; plateau: \SI{40}{s}; ramp down: \SI{10}{s}) with the maximal amplitudes ranging from \SIrange[range-phrase=\,\,to\,\,,range-units = single]{5}{10}{nA} (CoV of the common noise: \SI{20}{\%}, CoV of the independent noise: \SI{10}{\%}), we obtained a mean active pool of 89 MUs (standard deviation: 35 MUs). We added coloured Gaussian noise to obtain signal-to-noise ratios (SNR) ranging from \SIrange[range-phrase=\,--\,,range-units = single]{10}{30}{dB}. 
To focus on the subset of detectable MUs, we only considered MUs that had an amplitude larger than 3 and SEP larger than 0.5 at an extension factor $R=16$.
We found that all features (spike amplitudes, relative peak separation, and MUAP similarity) quickly approached a plateau for an extension factor in the range of 8 to 12 {(Figure~\ref{fig:extension_factor}A-C)}.
For $R=8$, the spike amplitudes were \SI{83.4}{\%}, \SI{84.1}{\%} and \SI{90.1}{\%} of the maximum values (i.e., at $R=24$) for the 10th, 50th and 90th percentile, respectively.
Considering the relative peak separation, an extension factor of 2 yielded \SI{0.4}{\%} (10th percentile), \SI{60.3}{\%} (50th percentile) and \SI{81.7}{\%} (90th percentile) of the values at $R=24$. 
Lastly, by increasing the extension factor from 8 to 24, the cosine similarity of the most similar extended and whitened MUAP decreased from 0.25 to 0.20 (10th percentile), 0.37 to 0.31 (50th percentile), and 0.60 to 0.50 (90th percentile).

\subsection{Source quality estimation}\label{sec:quality_source_metric}
\begin{figure}[ht!]
\centering
\includegraphics[width=\textwidth]{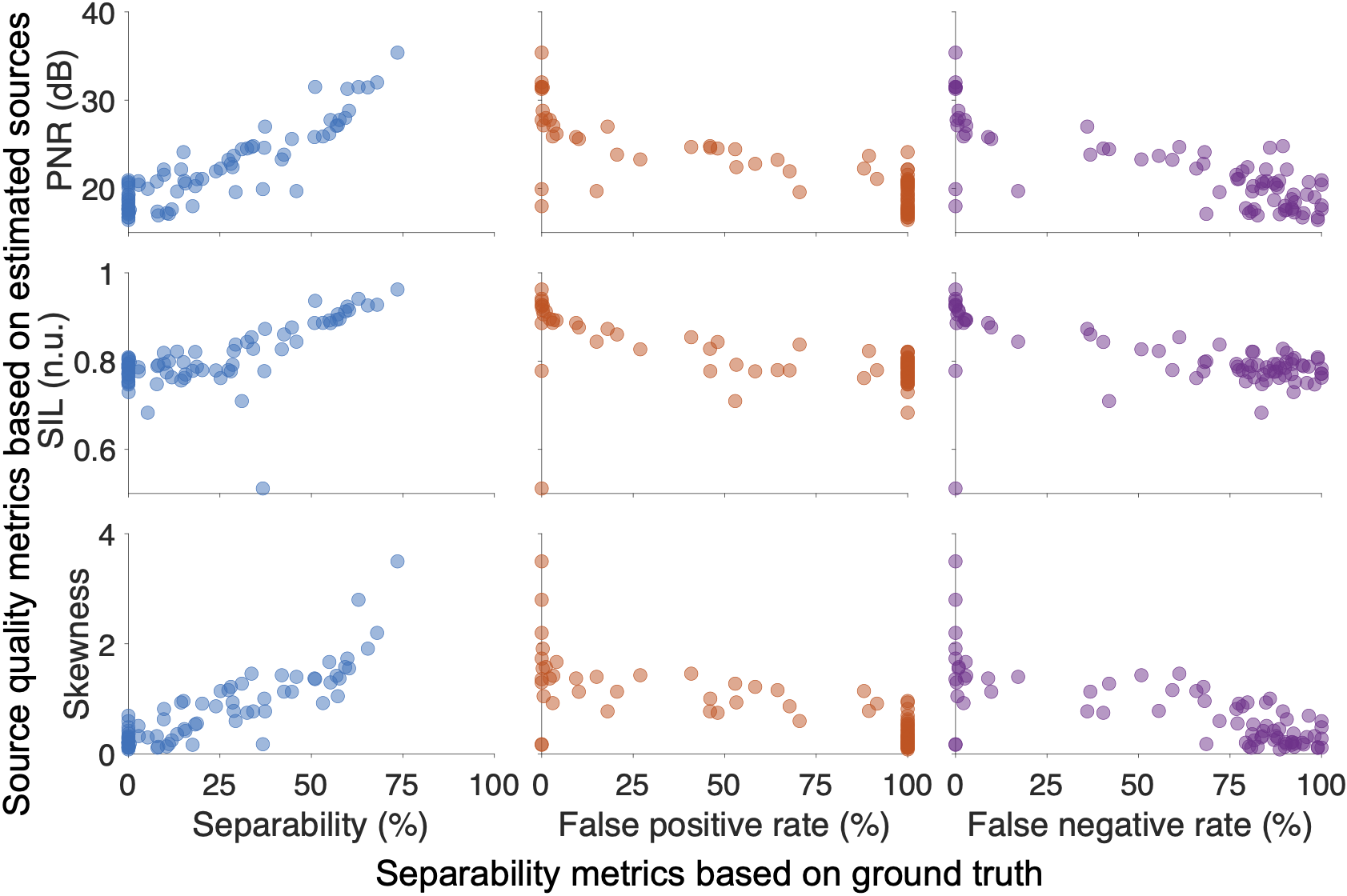}
\caption{Investigating the association of separability, false positive and false negative rates with source quality metrics: Pulse-to-Noise Ratio (PNR), Silhouette (SIL) value, and skewness. 
}
\label{fig:quality_source_metric}
\end{figure}

To study the association of source quality metrics, i.e., SIL, PNR and skewness, and the decomposition performance, we generated 64-channel surface EMG signals using a mean trapezoid drive at \SI{7}{nA} (CoV of the independent noise: \SI{5}{\%}, CoV of the common noise: \SI{20}{\%}, SNR: \SI{20}{dB}), resulting in a pool of 76 MUs. 
After reconstructing each source using the ground truth MUAP, we calculated the SIL, PNR, skewness, and finally, the separability as well as the false positive and negative rates {(Figure~\ref{fig:quality_source_metric})}.
PNR and SIL values have a limited effective range, i.e., \SIrange[range-phrase=\,--\,,range-units = single]{16}{35}{dB} and 0.51-0.96, in contrast to the skewness (0.1-3.5) and separability metric (\SIrange[range-phrase=\,--\,,range-units = single]{0}{74}{\%}). 
There was a general tendency that higher PNR and SIL values were related to a higher degree of separability and lower false positive and negative rates. On the other hand, SIL values around 0.85-0.90 resulted in a wide separability range (\SIrange[range-phrase=\,\,to\,\,,range-units = single]{0}{50}{\%}). PNR and skewness provided similar variations. It can also be seen that for relatively small PNR (\SI{18}{dB} and \SI{20}{dB}) and SIL (0.51 and 0.78) values, two MUs can still be separable with no false positive or negative rate. However, these were the two latest recruited MUs with few spikes, and it may be difficult to separate them unless a projection vector has already been trained.
The correlation between the separability metric and the PNR, SIL, and skewness was 0.90, 0.69, and 0.87, respectively ($p<0.001$). Moreover, the correlation between skewness and PNR was 0.91, and the correlation between skewness and SIL was 0.69 ($p<0.001$). Using kurtosis, the correlation with PNR and SIL was 0.92 and 0.70 ($p<0.001$). However, considering only SIL values larger or equal to 0.9, the correlation with skewness and kurtosis increases to 0.87 ($p<0.01$) and 0.93 ($p<0.001$).

\subsection{Population-based MU identification analysis}\label{sec:results_population}
After exemplarily showcasing factors affecting the performance of MU identification, we conducted a series of 200 in silico experiments to test the generalizability of the presented theory and observations.
All simulations used trapezoidal-shaped drives (ramp up: \SI{10}{ms}; plateau \SI{40}{ms}; ramp down \SI{10}{ms}). However, the peak amplitude (\SIrange[range-phrase=\,--\, ,range-units = single]{5}{10}{nA}), the CoV of the independent noise (\SIrange[range-phrase=\,--\, ,range-units = single]{0}{20}{\%}) and the CoV of the common noise (\SIrange[range-phrase=\,--\, ,range-units = single]{0}{60}{\%}) were randomly drawn from uniform distributions. 
This yielded an average number of 92 recruited MUs (standard deviation: 34) and spike independence coefficients ranging from 0.34 to 0.97.
To obtain EMG signals representing different virtual subjects, the selection of MUAPs and the selection of the EMG grid were randomized. Further, the additive noise was constructed to obtain SNRs between \SIrange[range-phrase=\,\,and\,\,]{10}{30}{dB} (drawn from a uniform distribution).
For the decompositions, we randomly selected an extension factor $R \in  [8, 20]$.

\begin{figure}[ht!]
\centering
\includegraphics[width=\textwidth]{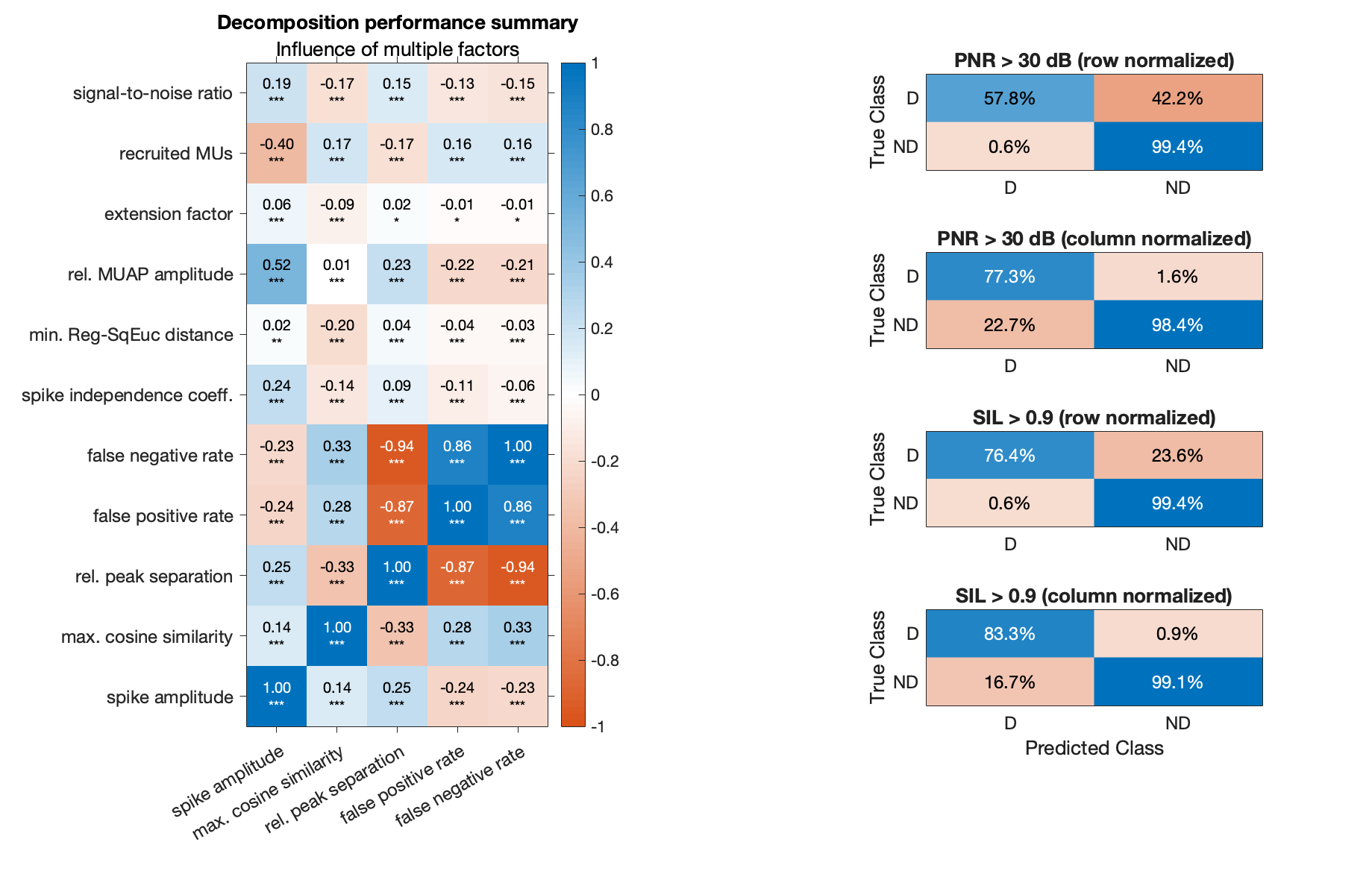}
\caption{Left: Correlation matrix, showing the influence of multiple factors on the decomposition performance. The significance of the correlations is indicated by symbols (*: $p<0.05$, **: $p<0.01$, ***: $p<0.001$). Right: Classification accuracy of Pulse-to-noise ratio (PNR) with threshold \SI{30}{dB} and silhouette (SIL) score with threshold 0.9 regarding detectable (D) and non-detectable (ND) motor neuron spike trains. Row normalization: Occurrence divided by the number of detectable / non-detectable sources in the ground truth data. Column normalization: Occurrence divided by the number of detectable / non-detectable motor unit (MU) sources given by the predicted labels.}
\label{fig:summary}
\end{figure}

Figure~\ref{fig:summary} (left) summarizes the results from the in silico trials, showing the linear correlation coefficients between a set of performance markers (expected spike amplitude, cosine similarity of the most similar extended and whitened MUAP, relative peak separation, false positive and false negative rate) and other factors previously shown relevant for decomposition performance (SNR, number of recruited MUs, extension factor, relative MUAP amplitude, Reg-SqEuc distance to the most similar MUAP and the spike independence coefficient).
For the external factors, the relative MUAP amplitude showed the strongest correlation with the false positive rate (correlation coefficient: $-0.22$) and the false negative rate (correlation coefficient: $-0.21$). 
Further, it is observed that reducing the SNR and increasing the number of recruited MUs had a similar effect, i.e., negatively affecting the identifiability of MUs. 
Violations of the independence assumption (causing lower values of the spike independence coefficient) reduced spike amplitudes (correlation coefficient: $0.24$) and increased the cosine similarity between the columns of the extended and whitened mixing matrix (correlation coefficient: $-0.14$). 
The latter was also connected to the Reg-SqEuc distance between the spatio-temporal MUAPs (correlation coefficient: $-0.20$).
For the range of extension factors, i.e., 8 to 20, $R$ showed only minor (yet significant) correlations with the decomposition performance. 
Decreasing the distance between MUAPs yielded higher false positive and negative rates.

The confusion maps (Figure~\ref{fig:summary}, right) summarize the classification accuracy of SIL (threshold value for classifying a MU as detectable: $\text{SIL}\geq 0.9$) and PNR (threshold value for classifying a MU as detectable: $\text{PNR}\geq$ \SI{30}{dB}) in distinguishing detectable and non-detectable sources.
As a ground truth, we considered spike trains detectable for which false positive and false negative rates were below \SI{10}{\%}.
The fraction of misclassified detectable sources was \SI{42}{\%} for PNR and \SI{24}{\%} for SIL.
Moreover, the fraction of sources falsely classified detectable was \SI{23}{\%} for PNR and \SI{17}{\%} for SIL.
We note that decreasing the thresholds of PNR and SIL to \SI{25}{dB} and 0.87, respectively, increased the true positive rates to \SI{90.4}{\%} (PNR) and \SI{92}{\%} (SIL). 
This comes at the cost of increasing the false positive rates to \SI{47}{\%} (PNR) and \SI{26}{\%} (SIL).
Further, note that using skewness and kurtosis to classify spikes (skewness threshold: $1.0$, kurtosis threshold: $7.5$) yielded true positive rates of \SI{83}{\%} (skewness) and \SI{91.7}{\%} (kurtosis) and false positive rates of \SI{35.0}{\%} (skewness) and \SI{49.6}{\%} (kurtosis).

\section{Discussion}
The theory and simulations presented in this study provide new insights into the existence and identifiability of inverse solutions when using convolutive BSS to identify spiking motor neuron activity, taking several factors into account. 
These insights have various practical implications when applying MU identification algorithms to specific signals, research questions, and clinical applications, which we will discuss in the following.

\subsection{Role of motor unit responses}\label{sec:discussion_muap}
The uniqueness of MUAPs is considered critical for decomposing surface EMG signals into individual MUs. For example,
{\citet{Farina2008}} introduced an empirical condition stating that an MU is identifiable if the regularized squared Euclidean distance (in the cited paper referred to as energy similarity) compared to all other MUs is greater than \SI{5}{\%}.
This study updates our understanding of the relevance of MUAP uniqueness by showing that the identifiability of MUs also depends on MUAP amplitude, noise and source statistics (see Sections~\ref{sec:results_similar_muaps} and \ref{sec:results_independence}).
This observation is closely related to the whitening transformation, which aims to orthogonalize the columns of the mixing matrix (see Section~\ref{sec:inverse_preconditioning}).
In the case of dependent MUAPs, the inversion of the mixing matrix is ill-conditioned and hence, the orthogonalization is only partially achieved for the most dominant (i.e., height amplitude) MUAPs.
This result is underlined by findings from previous studies.
In {\citet{Klotz2023}}, it was shown that when measuring magnetic muscle signals, the uniqueness of MU responses does not depend on the depth of MU territories. 
However, due to the signal decay over distance, the identifiability of MUs still shows a strong dependency on the depth of the MU territories.
Moreover, in {\citet{Caillet2023}}, it was shown that although increasing the density of surface electrode grids enhances the uniqueness of MUAPs, it is still hardly possible to detect low-threshold MUs. This is due to the (physiological) noise level.

Another key assumption in MU identification is the stationarity of MUAPs {\citep{Goodlich2023}}. In experimental surface EMG studies, one can (approximately) achieve stationarity using isometric and non-fatiguing exercise protocols. 
Nevertheless, (mild) violations of the stationarity assumption are unavoidable due to subtle movements, adaptivity of muscle fibre action potentials, sub-optimal fixation of subjects, etc. 
In this study, we show that convolutive BSS-based MU identification is robust against modest violations of the stationarity assumption. There is some wiggle room for non-stationary MUAPs, depending on the relative MUAP amplitude. Interestingly, while the overall probability of detecting MUs decreases with the relative MUAP amplitude, there is a tendency that MUs with high amplitude MUAPs are more sensitive to violations of the stationarity assumption (Table~\ref{tab:non-stationary}).
This observation agrees with recent findings from ultrasound-based decomposition studies {\citep{Lubel2024, Rohlen2025}}, where stationarity is not guaranteed even for isometric contractions.
Moreover, it has been shown that MU identification is sometimes feasible for (typically rather slow) dynamic contractions \citep[][]{Oliveira2021,Kramberger2021, Yokoyama2021,Guerra2024,Clarke2024,Glaser2018,Chen2020}.
Although MU identification during slow dynamic contractions, i.e., where the stationarity violation is modest (at least within limited temporal windows), is possible, the problem of decomposing surface EMG signals under arbitrary non-stationarities is still unresolved and the subject of ongoing research. 
Robustness to violations of the stationarity assumption is also important for MU tracking across recordings and sessions. 
In contrast to previous studies that used correlation for assessing MUAP similarity, we used a regularized squared Euclidean distance metric because correlation provides limited amplitude information and thus may result in false positives, particularly due to the low-pass filtering effect of the volume conductor.

Convolutive BSS methods can theoretically detect doublets (see Supplementary Figure~C2). 
In contrast to reports from previous studies {\citep{Chen2015}}, the duration of MUAPs does not need to be shorter than the inter-spike interval. 
This finding agrees with recent findings from the decomposition of ultrasound-based images {\citep{Lubel2024, Rohlen2025}}. 
For slow-twitch MUs, the mechanical impulse response after a neural discharge (MU twitch) can be \SI{300}{\milli\second} or more {\citep{McNulty2000, Raikova2008}}. 
This impulse response is about ten-fold longer than the duration of a surface EMG-based MUAP and still enables the robust identification of motor neuron spike trains. We did not vary the MUAP duration relative to the inter-spike interval. However, it is not the duration per se that is important; it is the relative duration of the two that matters. That is, instead of prolonging the MUAP duration artificially and non-physiologically, we focused on the doublets to test the relative duration of the MUAP and the inter-spike interval. The only requirement for detecting doublets is that two consecutive spikes are separated by approximately \SI{1}{\milli\second} and have nearly stationary MUAPs, which is important since the two consecutive MUAPs are usually not identical \citep{Piotrkiewicz2013}. 
Thus, the feasibility of detecting doublets needs to be verified experimentally.

\subsection{Relevance of independent sources}
Although MU sources are commonly assumed to be independent \citep[e.g.,][]{Merletti2008, Chen2015, Meng2022, Zheng2023}, ICA-based MU identification methods can identify theoretically dependent MU spike trains. 
Due to sparsity, MU spike trains often exhibit weak practical dependence (e.g., low mutual information).
Yet, the in silico experiments show that the expected MU yield decreases with increasing source correlation, where primarily, MUAPs with a lower norm become non-detectable. 
It is important to stress that the model is fitted to the observed data based on a linear superposition of the sources, the shape of the prior, and the joint prior being factorial (i.e., statistical independence). 
The objective function is constructed to favour projection vectors $\mathbf{w}_k$ that satisfy all three assumptions.
If any of the assumptions are incorrect, solving the given optimization problem still yields projection vectors that satisfy the other assumptions as well as possible.
Interestingly, it is described in the literature that ICA is closely related to sparse coding, which aims to find sparse representations of data {\citep{Hyvarinen1998b}}. 
For example, by using a sparse prior, an ICA model based on a fixed-point iteration algorithm has been used for learning the weights of a sparse coding basis in image denoising {\citep{Hyvarinen1998a}}. 
Similarly, the sparse spike trains, even if not truly independent, can be solutions to the posed optimization problem (because they are non-Gaussian).
It is acknowledged in the fundamental ICA literature that whitening solves half of the given BSS problem {\citep{Hyvarinen2000}}.
The effectiveness of the whitening depends on the mixing matrix itself and the covariance matrix of the extended sources (see Equation~\ref{eqn:covariance_decomposition}). 
Together, these factors determine the degree of ill-conditioning of the decomposition problem.
Moreover, the in silico experiments show that the norm of an extended and whitened MUAP decreases when the source synchronization increases, underscoring the relevance of whitening in explaining the performance decrease related to synchronized sources.

\subsection{Algorithm optimization and development}
The presented theory and simulations guide the optimization of convolutive BSS-based MU identification algorithms. 
It is shown that an optimal objective function is a compromise between enhancing the contrast between different sources and perturbations, i.e., noise and outliers (see Figure~\ref{fig:effect_of_exponent}). 
While this was recently shown empirically using a computationally expensive genetic algorithm {\citep{Grison2024, Grison2025}}, the theory presented in Section~\ref{sec:identifiability} can serve as a basis for a knowledge-based objective function selection.
Empirical observations suggest that a fixed exponent of 3 is a reasonable choice for most cases \citep{Negro2016,Grison2025}. From Figure~\ref{fig:effect_of_exponent}C it can be seen that effect of the non-linearity around the optimum is rather modest, underscoring the appropriateness of this hyperparameter choice.
 
In contrast to the existing literature, it is shown that the optimal extension factor $R$ is not explicitly related to the MUAP duration (see Section~\ref{sec:results_extension}). 
In the literature, it is reported that increasing the extension factor $R$ improves the ratio between observations and unknowns. Thus, the selection of the extension factor is guided by (semi-)analytical rules, e.g., $R \geq (N/M)\times L$ {\citep{Negro2016}}. 
This is only true if $M>N$, as each delayed copy of the observations adds one delayed source (per active MU).
However, even if $N>M$, increasing the extension factor $R$ enhances the spike amplitudes and reduces the linear dependency of the extended and whitened MUAPs.
Thereby, note that non-detectable MUs can always be treated as part of the noise.
Selecting $R=16$ has empirically been shown to yield the highest number of decomposed MUs \citep{Negro2016}.
Thus, many studies in the literature using convolutive BSS methods have used extension factors in that order of magnitude \citep[e.g.,][]{Avrillon2024, Lubel2024, Rohlen2025}; regardless of the contraction levels, the underlying signal (EMG or ultrasound) and regardless of the number of active MUs or the MU impulse response duration. 
Given the data presented in Section~\ref{sec:results_extension} (with an average active pool of $N=89$ MUs, $M=64$ channels, and $L=101$ temporal samples per MUAP), the empirical guideline is $R \geq (N/M) \times L=140$, which is considerably larger than the onset of the plateau (approximately $R=8$). 
To be more restrictive, the main energy of the MUAPs is concentrated in a window of 25 samples. Using $L=25$ samples yields $(N/M) \times L=34$, which is still about 4 times larger than $R=8$. Note that other empirical rules exist regarding the selection of the extension factor. 
Similar to the empirical rule discussed above, they can be considered to have a solid safety margin \citep[e.g.,][]{Farina2016a}.

A common issue in decomposition algorithms is the repeated convergence to the same source (or delayed versions of it).
There are two main approaches to mitigate this problem, i.e., (i) sub-space projection (also referred to as source deflation) and (ii) subtracting the already decomposed parts of the signal (also referred to as peel-off). 
The theory presented in Section~\ref{sec:existence} allows us to comprehensively understand the effect of both approaches. 
For source deflation, the MU filter is projected in a subspace orthogonal to previously identified MU filters. 
Accordingly, the cosine similarity (with respect to these sources) becomes zero, and hence, it effectively prevents the convergence to the same source.
Nevertheless, it is still possible to find the delayed copies of this source.
Moreover, source deflation comes with the cost of reducing the cosine similarity term for the source of interest (and thus the expected spike amplitude). 
Hence, it might limit the identifiability of low-amplitude MUs with low separation from the noise. 
On the other hand, as the peel-off approach subtracts the entire contribution from already decomposed MUs, it also prevents convergence to delayed copies of identified sources. 
Further, the expected spike amplitudes remain unaffected. 
However, estimating the signal contribution of already decomposed MUs yields errors that enhance the projected noise.
Thus, the advantages of a peel-off strategy can only be capitalized when a high signal-to-noise ratio can be guaranteed. 

We also want to stress that the further development of decomposition algorithms is facilitated by the support of open science principles.  
There is a particular need for open-source codes \citep[e.g.,][]{Avrillon2024, Valli2024, Formento2021, Grison2025} and community-accepted datasets for benchmarking.

\subsection{Source quality metrics}
Estimated MU spike trains are imperfect (see Section~\ref{sec:existence}). 
In detail, linear dependencies of the mixing matrix's columns cause cross-projection errors manifested as background peaks. 
Due to the (random) superposition of active MUs (or potentially any other noise source), background spikes can even become considerably higher than the expected cross-projection errors (see Section~\ref{sec:results_accuracy}). 
While it is known that decomposition performance critically depends on the measurement noise {\citep[e.g.,][]{Avrillon2024b}}, physiological noise is unavoidable.
Hence, an essential part of MU identification is quantifying the uncertainty associated with estimated spike trains.

For the decomposition of experimentally acquired signals, the ground truth is unknown, making the development of proper source quality metrics challenging.
This work demonstrates that widely used source quality metrics like PNR and SIL provide only limited insight into the goodness of estimated spike trains. 
SIL is a highly non-linear measure of the separation between the peak cluster and the background peak cluster in terms of the standard deviation of the peak cluster (e.g., $\text{SIL}\approx 0.9$ indicates a separation by 3 standard deviations, and $\text{SIL}\approx 0.86$ indicates a separation by 2.5 standard deviations).
Hence, subtle changes in the SIL score can indicate strongly deviating uncertainties. 
Moreover, the variability of the background peak cluster is not considered. 
PNR essentially compares the spike amplitudes against the noise and background peaks. 
The variability of the spike amplitudes is not considered. 
As the presented theory clearly shows the role of spike variability on the robustness of the decomposition problem, SIL is considered superior to PNR.
This superiority is underscored by the presented simulations, whereby SIL-based source classification achieved a better performance than PNR-based source classification (see Section~\ref{sec:results_population}).
We have further demonstrated that PNR is closely related to evaluating the objective function of a given source. 
Indeed, PNR is highly correlated with the sample skewness and kurtosis of the source, which should perform similarly as a source quality metric and is, e.g., done in calcium imaging {\citep{Mukamel2009}}. In fact, SIL is also highly correlated with skewness and kurtosis when only considering sources with a SIL $\geq 0.9$. 
Given currently available source quality metrics, we recommend using SIL due to its superiority over other existing metrics (see Figure~\ref{fig:summary} and Section~\ref{sec:results_population}).

Due to the issues of existing signal-based uncertainty measures, sometimes physiological criteria are used as source quality metrics, e.g., based on the coefficient of variation of the inter-spike intervals. 
When using experimental conditions with known physiological behaviour, this is reasonable for validation purposes \citep{Enoka2019}, but it is inappropriate for studying novel phenomena.
Hence, advancing unsupervised source quality metrics remains a key challenge to improve fully automated and non-invasive MU identification methods.

\subsection{Relevance for clinical applications}
Considering the effects of MU synchronization, MUAP similarity and non-stationarity are crucial for translating motor neuron identification algorithms to clinical applications.
The results indicate that increased spike train correlations decrease the expected MU yield.
This may be important for several patient groups, e.g., with essential tremor \citep{Hellwig2003}.
Yet, the MUAPs with larger norms may still be detected, as demonstrated in \citet{Holobar2012}. 
Further, it is known that muscle fatigue causes signal non-stationarities.
Abnormal fatigability and exercise intolerance are common in a wide variety of human diseases, ranging from primary motor neuron diseases such as Spinal Muscular Atrophy (SMA) \citep{Montes2021}, and degenerative bodily states such as heart failure \citep{DelBuono2019}, cancer \citep{Hussey2022} and ageing \citep{Wu2021} to post-viral syndromes, e.g., Chronic Fatigue Syndrome \citep{Gherardi2019} and Long-Covid \citep{Silva2022}.
We have shown that there is some wiggle room for non-stationarities. Yet, applying ICA-based MU identification methods to such patient groups might require further development of algorithms capable of handling non-stationarities.
Reinnervation in response to denervation may lead to larger MUAPs \citep{Stalberg2011}, increasing the detectability for a few MUs at the cost of a smaller MU yield. On the other hand, some patient groups of interest may have a thick fat layer, leading to a longer distance between electrodes and the MU sources, which results in lower MUAP amplitudes and less SNR, reducing the ability to detect MUs. Other patient groups having atrophy will lead to a closer distance between electrodes and MU sources, leading to reduced filtering of the MUAPs due to the volume conductor, but have the risk of crosstalk from nearby muscles. 

\subsection{Limitations}
This study has several limitations. The study relies solely on computer simulations, which are limited by the constraints of the computational model used. However, they are effective in demonstrating theoretical findings and constructing scenarios such as perturbing an identifiable MU by, for example, making another MU more or less similar or modulating the synchronization of the active MU pool. It is extremely challenging, if not impossible, to do this by means of experiments.

For simulating EMG signals, we used a leaky integrate and fire model and MUAPs extracted from experimental data.
While this provides a good trade-off between the level of biophysical accuracy and complexity, this modeling approach has many shortcomings. Although the selected MUAPs were extracted from contraction ranging from 10 to \SI{80}{\%} MVC, the EMG model may create a bias because it is based on MUAPs that are decomposable.
However, we focus on identifiable MUs and perturb them to study what happens to the identifiability.
The motor neuron model captures important aspects such as the size principle \citep{Henneman1954} and realistic firing rates \citep{Fuglevand1993} while processing input currents closely matching our current understanding of motor control \citep[e.g.,][]{Heckman2012, Farina2014, Hug2025}.
However, the model does not resolve key aspects of motor neuron physiology like neuromodulation \citep[e.g.,][]{Heckman2008, Mesquita2024} or doublets \citep{Christie2006,Mrowczynski2015}.
In this study, we simulated doublets by adding extra spikes to the computed spike trains.
Explicitly modelling such behaviours may require a biophysically more detailed motor neuron model \citep[e.g.,][]{Cisi2008, Powers2012, Schmid2024}. 
However, this detailed modelling is beyond the scope of this work. 
Further, using experimentally extracted MUAPs, along with a simple MUAP similarity generator using interpolation, does not allow us to directly relate the simulation results and the structure and function of the muscle, requiring the use of a biophysical skeletal muscle model \citep[e.g.,][]{Farina2001, Lowery2002, Klotz2020}. Coloured Gaussian noise may be a simplified noise source. However, given the fact that non-detectable MUs can be treated as part of the noise, and in the simulated datasets, the fraction of non-detectable MUs is typically much higher than the fraction of detectable MUs, the structure of the effective noise is more complex than uncorrelated Gaussian noise. Further, it still allows for studying the effect of the signal-to-noise ratio.

The decompositions presented in this work make use of the known mixing model to estimate the MU spike trains \citep[][]{Klotz2023}. 
We show that for robust decompositions, (local) optima of the objective function correspond to the extended and whitened MUAPs.
Nevertheless, solving a convolutive BSS problem might give different results. 
Depending on the selected parameters, some sources might not be part of the mathematical solution space or might be missed by the optimization algorithm due to the complexity of the optimization landscape.
Hence, the results represent an upper-bound performance for estimating MU spike trains using convolutive BSS. 

\subsection{Conclusion}
In conclusion, we revisited the EMG-based convolutive BSS model, derived theoretical results, and verified them using in silico experiments, leading to an improved understanding of methods used for non-invasive motor neuron identification. 
These findings will play an important role in advancing MU identification algorithms, translation to signals other than EMG and applicability in various groups of patients.

\section*{Acknowledgements}
TK is supported by the Deutsche Forschungsgemeinschaft (DFG, German Research Foundation) through the priority program SPP 2311 (Grant ID: 548605919) and the European Research Council (ERC) through the ERC-AdG ``qMOTION'' (Grant ID: 101055186). RR is supported by the Swedish Brain Foundation (Grant ID: PS2022-0021), the Swedish Research Council (Grant ID: 2023-06464), and the Promobilia Foundation (Grant ID: A23161). 

\section*{Data availability statement}
All data and codes needed to replicate the results presented in this manuscript are available on GitHub (URL: \url{https://github.com/klotz-t/MUDecompositionTests}, Version 1.1.0: \url{https://doi.org/10.5281/zenodo.16794779}).

\appendix
\renewcommand{\figurename}{Supplementary Figure}
\setcounter{figure}{0}

\section{Supplementary theory}
\renewcommand{\theequation}{A.\arabic{equation}}
\setcounter{equation}{0}
\subsection{Non-whitened spike estimation}\label{appendix:ckc_spike_estimation}
The spike estimation can also be conducted in the non-whitened space {\citep[cf. e.g.,][]{Holobar2007a, Farina2016a}}, yielding
\begin{equation}\label{eqn:spike_estimation_non_whitened}
\begin{split}
   \widehat{\tilde{s}}_k(t) \ &= \mathbf{w}_k^T \, \tilde{\mathbf{z}}(t) \ = \ (\mathbf{V} \,\mathbf{r}_k)^T (\mathbf{V\mathbf{\tilde{x}}}(t)) \\ 
   &= \ \mathbf{r}_k^T \mathbf{V}^T \mathbf{V}\mathbf{\tilde{x}}(t) \ = \ \mathbf{r}_k^T \mathbf{C}_{\tilde{x}\tilde{x}}^{-1} \tilde{\mathbf{x}}(t) \ . 
\end{split}
\end{equation}
Therein, $\mathbf{r}_k$ is the non-whitened projection vector.
Further, the covariance matrix of the whitened data is the identity matrix by definition, i.e., $\mathbf{V} \mathbf{C}_{\tilde{x}\tilde{x}} \mathbf{V}^T = \mathbf{I}$, which, under the assumption that $\mathbf{V}$ is invertible, is used to obtain $\mathbf{C}_{\tilde{x}\tilde{x}}^{-1} = \mathbf{V}^T\mathbf{V}$.
Analogous to the whitening transformation, the (pseudo)inverse of the extended signal's covariance matrix uncorrelates the extended observations. 
Whitening the extended observations is often numerically more robust than computing the (pseudo)inverse of $\mathbf{C}_{\tilde{x}\tilde{x}}$.

\subsection{Extended and whitened mixing matrix}\label{appendix:ext_and_white_mixing_matrix}
The columns of the extended and whitened mixing matrix $\mathbf{\tilde{h}}_{ul}$ are given by
\begin{equation}
\label{eqn:MUAP_filter}
\begin{split}
    \mathbf{\tilde{h}}_{ul} = \mathbf{V}\mathbf{\tilde{H}}_{:,u}(:,l) \ = \ \mathbf{V}\Big[&h_{1u}(l+R-1), \, ... \, , h_{1u}(l) \, , \, ... \, , \\
    &\qquad h_{Mu}(l+R-1), \, ... \, , h_{Mu}(l)\Big]^T \ ,
\end{split}    
\end{equation}
with $0 \leq l \leq L + R - 1$.

\subsection{Taylor series of the expected objective function value}\label{appendix:taylor_series_cf}
The expected value of the contrast function applied to the estimated sources $\mathbb{E}(G(s))$ can be approximated by a Taylor series expansion of $G(s)$ around the mean $\mu$:
\begin{equation}
\label{eqn:Loss_Taylor}
\begin{split}
    \mathbb{E}(G(s)) \ &= \ G(\mu) \, + \,  \sum_{n=1}^{\infty} \mathbb{E}\left(\frac{G^{(n)}(\mu)}{n!} (s-\mu)^n \right) \\
    &= \ G(\mu) \, + \,  \sum_{n=1}^{\infty} G^{(n)}(\mu) \,  \mathbb{E}\left(\frac{(s-\mu)^n}{n!} \right) \\
    &= \ G(\mu) \, + \, \frac{G''(\mu)}{2} \, \sigma^2 + \, \frac{G^{(3)}(\mu)}{6} \, \gamma_1 \sigma^3 \, + \, \mathcal{O}(s^4) \ .
\end{split}
\end{equation}
Therein, the first-order term vanishes as $\mathbb{E}(s-\mu)=0$ and we substitute the definition of the variance $\sigma^2 = \mathbb{E}[(s-\mu)^2]$ and skewness $\gamma_1 = \mathbb{E}[(s-\mu)^3]/\sigma^3$.

\subsection{Single spike solutions}\label{appendix:single_spike}
A common phenomenon in ICA-based source estimates is the identification of non-physiological single-spike sources. 
In Section~\ref{sec:identifiability}, it was shown that the objective function rewards outliers, see Equation~\eqref{eqn:approx_loss}. 
Here, we showcase the extreme case, where the outlier spike dominates the total objective function.
To do so, we chose a projection vector $\mathbf{w}_k$ that is perfectly aligned with a single time frame of the extended observations $\mathbf{\tilde{z}}(t_{i_k})$ with $t_{i_k} \in \mathcal{S}_k$ and $\mathcal{S}_k$ being the set of spikes of MU $k$ (with delay $l$).
If  $\mathbf{\tilde{z}}(t_{i_k})$ is an outlier only weakly correlated with $\mathbf{\tilde{h}}_{kl}$, the optimum of $L(\mathbf{w}_k)$ can correspond to a projection vector highlighting this single outlier spike.
Mathematically, this can be shown if we only consider true spikes, i.e., $t \in \mathcal{S}_k$, and restrict $a>1$, i.e., yielding a convex function. Making use of Jensen's inequality, we obtain the following condition:
\begin{equation}\label{eqn:noise_inequality}
    \mathbb{E}\Big[\sum_{t\in\mathcal{S}_k}  G\Big(\mathbf{w}_k^T \mathbf{\tilde{z}}(t)  \Big)\Big] \ \geq \  N_{\mathcal{S}_k} \cdot G\left(S_\mathrm{cos}^{k,kl} ||\mathbf{\tilde{h}}_{kl}||\right)  > \    G \, \Big(  \underset{t \in \mathcal{S}_k}{\operatorname{arg \, max}} \ ||\mathbf{\tilde{z}}(t)||
    \Big) \ .
\end{equation}
Hence, a true source is only identifiable if there exists a $\mathbf{w}_k$ such that \eqref{eqn:noise_inequality} holds. 
It can be seen that the identifiability depends on the number of spikes, the selected objective function and the ratio between the amplitudes of the outlier and $S_\mathrm{cos}^{k,kl} ||\mathbf{\tilde{h}}_{kl}||$. 

\subsection{Source quality metrics}\label{appendix:source_quality_metrics}
The SIL score for the $k$th MU source is defined as 
\begin{equation}\label{eqn:def_sil}
    \mathrm{SIL}^k \ = \ \frac{\sum\limits_{i \in \mathcal{S}_\mathrm{p}} \Big(\varphi^k(t_i) - \mu_\mathrm{p}\Big)^2 - \sum\limits_{i \in \mathcal{S}_\mathrm{p}} \Big(\varphi^k(t_i) - \mu_\mathrm{n}\Big)^2}{\max \left\{\sum\limits_{i \in \mathcal{S}_\mathrm{p}} \Big(\varphi^k(t_i) - \mu_\mathrm{p}\Big)^2, \sum\limits_{i \in \mathcal{S}_\mathrm{p}} \Big(\varphi^k(t_i) - \mu_\mathrm{n}\Big)^2 \right\}}  \ \ .
\end{equation}
Therein, $\mathcal{S}_\mathrm{p}:=\{t^1_\mathrm{p}, ..., t_\mathrm{p}^{N_\mathrm{p}}\}$ is the set of discharge times of MU $k$ (cf. Equation~\ref{eqn:convBSSdirac}), with an expected spike amplitude of $\mu_\mathrm{p}=\mathbb{E}[\varphi(t_i)]$. Further, $\mu_\mathrm{n}=\mathbb{E}[\varphi(t_j)]$ is the expected amplitude of spikes classified as false spikes.
From Equation~\eqref{eqn:separation_of_MUs}, it can be seen that MU identification is only possible if $\mu_\mathrm{p} > \mu_\mathrm{n}$.
Hence, without loss of generality, we can rewrite the SIL value in terms of the expected values
\begin{equation}\label{eqn:sil_alt}
    \mathrm{SIL}^k \ = \ \frac{\mathbb{E}\big[(\varphi^k(t_i) - \mu_\mathrm{p})^2\big] - \mathbb{E}\big[(\varphi^k(t_i) - \mu_\mathrm{n})^2\big]}{\mathbb{E}\big[(\varphi^k(t_i) - \mu_\mathrm{n})^2\big]} \ = \ \frac{(\mu_\mathrm{p}-\mu_\mathrm{n})^2}{\sigma_\mathrm{p}^2 + (\mu_\mathrm{p}-\mu_\mathrm{n})^2} \ \ ,
\end{equation}
where $\sigma_\mathrm{p}^2=\mathbb{E}[(\varphi(t_i)-\mu_\mathrm{p})^2]$ is the variance of spikes belonging to MU $k$.\\

The PNR for the $k$ MU source is defined as 
\begin{equation}\label{eqn:def_pnr}
    \mathrm{PNR}^k \ = \ 10 \cdot \log_{10} \left(\frac{\mathbb{E}[\varphi^k(t_i)^2]}{\mathbb{E}[\varphi^k(t_j)^2]} \right) \,
\end{equation}
where $t_j$ denotes all time frames not associated with the activity of MU $k$.
Further, note that in practice, $t_i$ often also includes samples in close neighbourhoods to the actual spike (e.g., $\pm1$ sample).
This is due to the fact that the width of the predicted spikes is typically larger than one sample.

\section{Supplementary simulations}
\renewcommand{\theequation}{B.\arabic{equation}}
\setcounter{equation}{0}
\subsection{Motor neuron parameters}\label{appendix:mn_params}
The motor neuron model parameters were based on experimental animal models {\citep{Caillet2022}}. We used the following parameters:
\begin{equation}
\begin{split}
R_j \ &= \ 1.68 \cdot 10^{-10} \cdot S_{j}^{-2.43},\\
\tau_j \ &= \ 7.9 \cdot 10^{-5} \cdot D_{j} \cdot R_j,\\
t_j^\mathrm{ref} \ &= \  0.2 \cdot 2.7 \cdot 10^{-8} \cdot D_{j}^{-1.51},\\
S_{j}\ &= \ 3.96 \cdot 10^{-4} \cdot I_{j}^{0.396},\\
I_{j}\ &= \ 3.85 \cdot 10^{-9} \cdot 9.1^{(j/N)^{1.1831}},
\end{split}
\end{equation}
where $R_j$, $\tau_j$, $t_j^\mathrm{ref}$, $S_{j}$, and $I_{j}$ denote the membrane resistance (unit: \SI{}{\ohm}), membrane time constant (seconds), refractory time (seconds), neuron surface area (unit. \SI{}{m^2}), and rheobase (ampere) for motor neuron $j$ ($j=1,\ldots,N$), with $N=300$ motor neurons. Moreover, $D_{j}$ denote the soma diameter for motor neuron $j$, which was set in a linear fashion from \SIrange[range-phrase=\,\,to\,\, ,range-units = single]{50}{100}{\micro\meter}. We set the time step to $\Delta t=$ \SI{0.1}{\milli\second}  (i.e., \SI{10}{kHz} sampling rate), the resting potential to $V_\mathrm{r}=$ \SI{-70}{\milli\volt}, the threshold potential to $V_{th}=$ \SI{-50}{\milli\volt}, duration $T=$ \SI{60}{\second}, and finally, the leakage and excitability
parameters $g_{j}^\mathrm{L}=g_{j}^\mathrm{E}$ linearly decreasing from 0.25 to 0.15 (smallest to largest).
The selected values provided realistic frequency-current behaviour for a large range of input currents, with the early recruited motor neurons showing frequency saturation (Supplementary Figure~C1).

\section{Supplementary data}
\renewcommand{\theequation}{B.\arabic{equation}}
\setcounter{equation}{0}
\subsection{Motor neuron parameters}\label{appendix:mn_params}
\begin{figure}[ht!]
\centering
\includegraphics[width=0.8\textwidth]{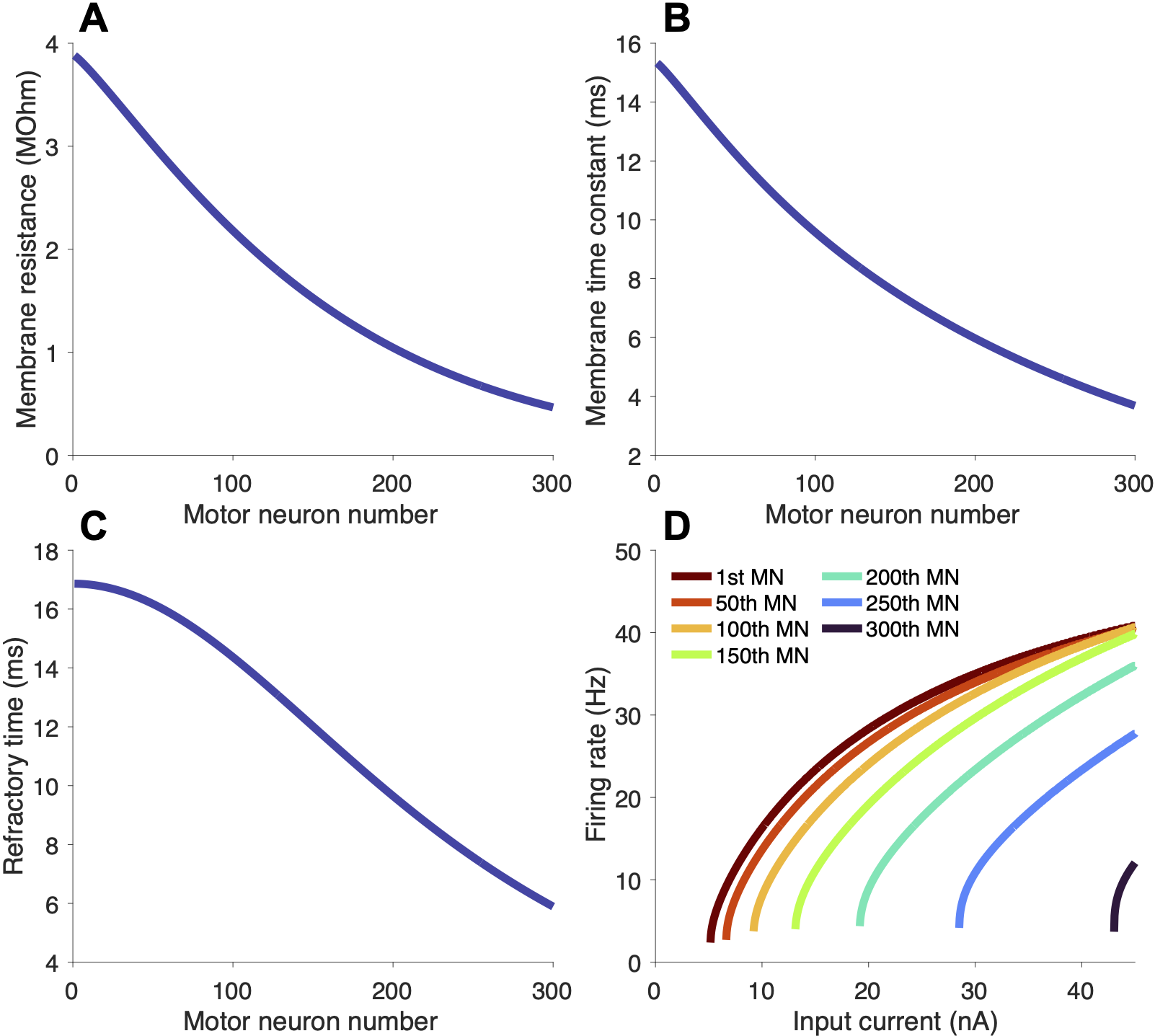}
\caption{Parameters of the leaky integrate and fire model depending on motor neuron (MN) size, i.e., the membrane resistances (\textbf{A}), membrane time constants (\textbf{B}), and refractory times (\textbf{C}). Firing rate for the exemplary selected motor neurons as a function of input current (\textbf{D}).}
\label{fig:LIF}
\end{figure}

Supplementary Figure~\ref{fig:LIF} summarizes the distribution of the motor neuron pool model parameter distribution, together with the current-frequency relation for a subset of exemplarily selected motor neurons.

\subsection{Doublets}\label{appendix:res_doublets}
\begin{figure}[h!]
\centering
\includegraphics[width=\textwidth]{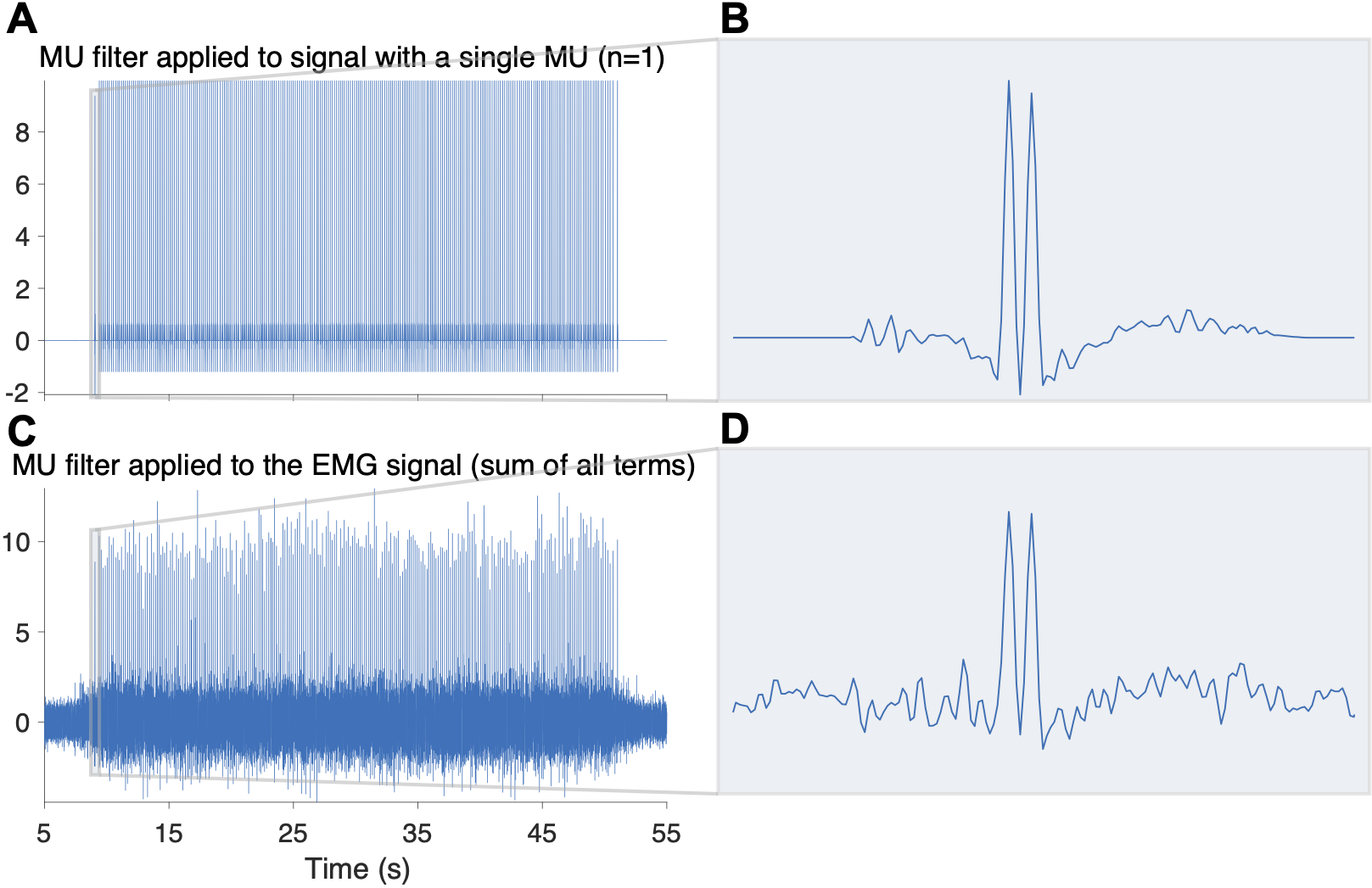}
\caption{Investigating the reconstruction accuracy of a motor unit (MU) with doublets (interspike-intervals: \SI{3}{ms}). The MU filter is applied to a noiseless EMG signal with only MU \#50 active (\textbf{A} and \textbf{B}) as well as the full EMG signal with additive coloured noise (\textbf{C} and \textbf{D}).}
\label{fig:doublets}
\end{figure}

We investigated the motor neuron identifiability in the presence of doublets, i.e., two consecutive discharges that are about \SIrange[range-phrase=\,\,to\,\, ,range-units = single]{2}{10}{ms} apart, which is important for rapid force generation. 
We generated surface EMG signals given an active pool of 78 MUs (trapezoid-shaped drive, maximal amplitude of \SI{7}{nA}, common noise with a CoV of \SI{20}{\%}, independent noise with a CoV of \SI{5}{\%} and an SNR of \SI{20}{dB}).
Then, we added an extra spike 3 ms after the first spike (corresponding to an instantaneous firing rate of \SI{333}{Hz}). We applied the MUAP-based projection vector to a noiseless EMG signal with only MU \#50 active and the full EMG signal with additive noise (Supplementary Figure~\ref{fig:doublets}). 
Although the spike reconstruction is imperfect, the resolution is good enough to separate the two spikes.

\subsection{Spike train dependencies}\label{appendix:common_spikes_norm}
\begin{figure}[h!]
\centering
\includegraphics[width=0.8\textwidth]{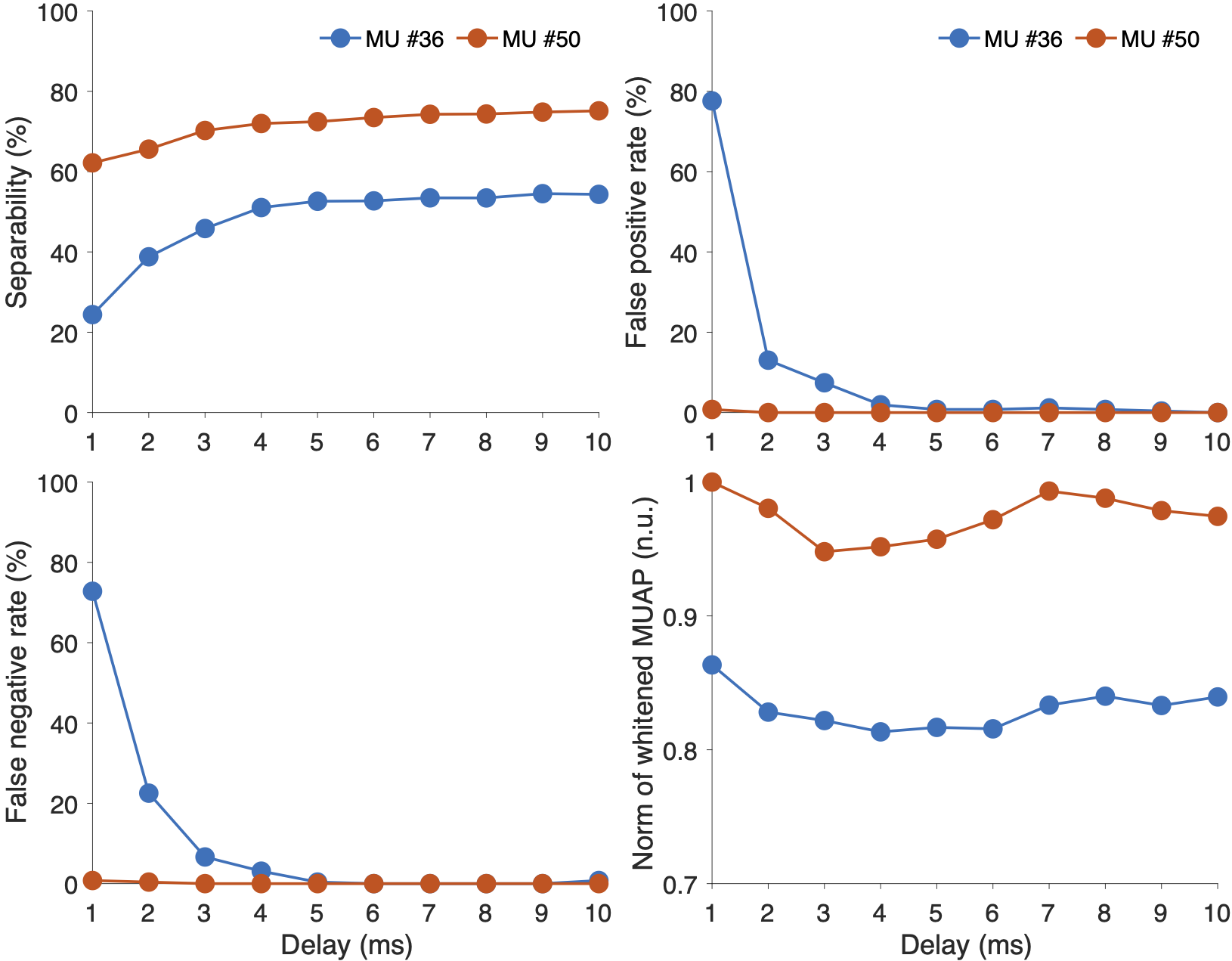}
\caption{Investigating the separability when two motor units (MUs) had identical spike trains but where one of the spike trains was slightly shifted (1 to 10 ms), inducing dependence between the two MUs.
}
\label{fig:delayed_spike_train}
\end{figure}

We also investigated motor neuron identification when two MUs (\#36 and \#50) had identical spike trains but one of the spike trains was slightly delayed (\SIrange[range-phrase=\,\,to\,\, ,range-units = single]{1}{10}{ms}). We found that for MUs with high amplitude MUAPs, it did not affect their separability {(Supplementary Figure~\ref{fig:delayed_spike_train})}. 
On the other hand, the MU with a lower MUAP norm was separable with minimal to no false positive and negative rates for about \SI{5}{ms} and more. For shorter delays, the false positive and negative rates increased rapidly.

\subsection{Non-stationary signals}\label{appendix:non-stationary}
\begin{figure}[hb!]
\centering
\includegraphics[width=\textwidth]{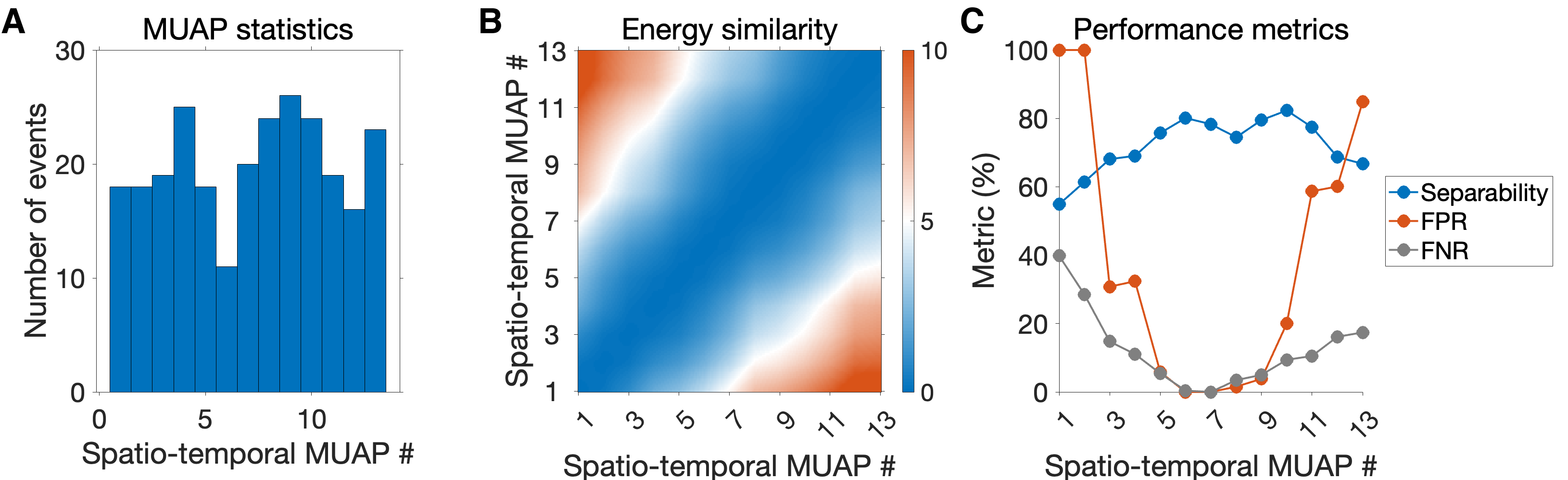}
\caption{Influence of signal non-stationarity on MU identification. Number of events per spatio-temporal MUAP (\textbf{A}). Regularized squared Euclidean (Reg-SqEuc) distance between all pairs of changing MUAPs (\textbf{B}). Performance metrics, i.e., separability, false positive rate (FPR) and false negative rate (FNR) when the projection vector is constructed from the individual 13 MUAP templates (\textbf{C}).}
\label{fig:changing_muaps}
\end{figure}

To investigate how non-stationarities, i.e., varying MUAP shapes between the discharges of the same motor neuron, affect the motor neuron identification, we simulated a trapezoidal-shaped contraction (ramp up: \SI{10}{s}; plateau: \SI{40}{s}; ramp down: \SI{10}{s}; max. drive: \SI{7}{nA}; CoV of the common noise: \SI{20}{\%}; CoV of the independent noise: \SI{5}{\%}) and assumed an SNR of \SI{20}{dB}.
Thereby, we randomly modulated the MU impulse responses {(Figure~\ref{fig:changing_muaps}A)} given 13 spatio-temporal MUAP templates (see Section~3.3).
The Reg-SqEuc distance of spatio-temporal MUAPs varied from close to \SI{0}{\%} to about \SI{11}{\%} {(Figure~\ref{fig:changing_muaps}B)}. 
By using the ground truth spatio-temporal MUAP for reconstructing the MU spike train, we found that using the first or last MUAPs could not separate the MU {(Figure~\ref{fig:changing_muaps}C)}. 
However, one can reconstruct the spike train using the MUAP templates \#6 to \#7. For these MUAP templates, the Reg-SqEuc distance with respect to all other MUAP templates is less than \SI{5}{\%} {(Figure~\ref{fig:changing_muaps}A)}. 
Note that these spatio-temporal MUAPs (\#6 and \#7) were highly similar to the average MUAP, i.e., the Reg-SqEuc distance is \SI{0.27}{\%} and \SI{0.31}{\%}, respectively.

\clearpage
\def\newblock{\ }%
\bibliographystyle{abbrvnat}      
\bibliography{bibliography}

\begin{thebibliography}{80}
\providecommand{\natexlab}[1]{#1}
\providecommand{\url}[1]{\texttt{#1}}
\expandafter\ifx\csname urlstyle\endcsname\relax
  \providecommand{\doi}[1]{doi: #1}\else
  \providecommand{\doi}{doi: \begingroup \urlstyle{rm}\Url}\fi

\bibitem[Avrillon(2023)]{Avrillon2023_data}
S.~Avrillon.
\newblock {Data for the preprint 'The decoding of extensive samples of motor
  units in human muscles reveals the rate coding of entire motoneuron pools'},
  11 2023.
\newblock URL \url{https://doi.org/10.6084/m9.figshare.24640944.v1}.

\bibitem[Avrillon et~al.(2024{\natexlab{a}})Avrillon, Hug, Baker, Gibbs, and
  Farina]{Avrillon2024b}
S.~Avrillon, F.~Hug, S.~N. Baker, C.~Gibbs, and D.~Farina.
\newblock Tutorial on {MUedit}: An open-source software for identifying and
  analysing the discharge timing of motor units from electromyographic signals.
\newblock \emph{Journal of Electromyography and Kinesiology}, 77:\penalty0
  102886, 2024{\natexlab{a}}.

\bibitem[Avrillon et~al.(2024{\natexlab{b}})Avrillon, Hug, Enoka, Caillet, and
  Farina]{Avrillon2024}
S.~Avrillon, F.~Hug, R.~Enoka, A.~H. Caillet, and D.~Farina.
\newblock The decoding of extensive samples of motor units in human muscles
  reveals the rate coding of entire motoneuron pools.
\newblock \emph{eLife}, 13, 2024{\natexlab{b}}.

\bibitem[Caillet et~al.(2022)Caillet, Phillips, Farina, and
  Modenese]{Caillet2022}
A.~H. Caillet, A.~T. Phillips, D.~Farina, and L.~Modenese.
\newblock Mathematical relationships between spinal motoneuron properties.
\newblock \emph{eLife}, 11:\penalty0 e76489, 2022.

\bibitem[Caillet et~al.(2023)Caillet, Avrillon, Kundu, Yu, Phillips, Modenese,
  and Farina]{Caillet2023}
A.~H. Caillet, S.~Avrillon, A.~Kundu, T.~Yu, A.~T. Phillips, L.~Modenese, and
  D.~Farina.
\newblock Larger and denser: an optimal design for surface grids of {EMG}
  electrodes to identify greater and more representative samples of motor
  units.
\newblock \emph{eNeuro}, 10\penalty0 (9), 2023.

\bibitem[Chen et~al.(2020)Chen, Ma, Sheng, Farina, and Zhu]{Chen2020}
C.~Chen, S.~Ma, X.~Sheng, D.~Farina, and X.~Zhu.
\newblock Adaptive real-time identification of motor unit discharges from
  non-stationary high-density surface electromyographic signals.
\newblock \emph{IEEE Transactions on Biomedical Engineering}, 67\penalty0
  (12):\penalty0 3501--3509, 2020.

\bibitem[Chen et~al.(2023)Chen, Ma, Sheng, and Zhu]{Chen2023}
C.~Chen, S.~Ma, X.~Sheng, and X.~Zhu.
\newblock A peel-off convolution kernel compensation method for surface
  electromyography decomposition.
\newblock \emph{Biomedical Signal Processing and Control}, 85:\penalty0 104897,
  2023.

\bibitem[Chen et~al.(2025)Chen, Li, and Xia]{Chen2025}
C.~Chen, D.~Li, and M.~Xia.
\newblock A motor unit action potential-based method for surface
  electromyography decomposition.
\newblock \emph{Journal of NeuroEngineering and Rehabilitation}, 22\penalty0
  (1):\penalty0 1--16, 2025.

\bibitem[Chen and Zhou(2015)]{Chen2015}
M.~Chen and P.~Zhou.
\newblock A novel framework based on {FastICA} for high density surface {EMG}
  decomposition.
\newblock \emph{IEEE Transactions on Neural Systems and Rehabilitation
  Engineering}, 24\penalty0 (1):\penalty0 117--127, 2015.

\bibitem[Chen and Zhou(2024)]{Chen2024}
M.~Chen and P.~Zhou.
\newblock {2CFastICA}: A novel method for high density surface {EMG}
  decomposition based on kernel constrained {FastICA} and correlation
  constrained {FastICA}.
\newblock \emph{IEEE Transactions on Neural Systems and Rehabilitation
  Engineering}, 2024.

\bibitem[Christie and Kamen(2006)]{Christie2006}
A.~Christie and G.~Kamen.
\newblock Doublet discharges in motoneurons of young and older adults.
\newblock \emph{Journal of Neurophysiology}, 95\penalty0 (5):\penalty0
  2787--2795, 2006.

\bibitem[Cisi and Kohn(2008)]{Cisi2008}
R.~R. Cisi and A.~F. Kohn.
\newblock Simulation system of spinal cord motor nuclei and associated nerves
  and muscles, in a web-based architecture.
\newblock \emph{Journal of computational neuroscience}, 25:\penalty0 520--542,
  2008.

\bibitem[Clarke et~al.(2024)Clarke, Grison, Guerra, Mamidanna, Ma, Muceli, and
  Farina]{Clarke2024}
A.~K. Clarke, A.~Grison, I.~M. Guerra, P.~Mamidanna, S.~Ma, S.~Muceli, and
  D.~Farina.
\newblock {HarmonICA}: Neural non-stationarity correction and source separation
  for motor neuron interfaces.
\newblock \emph{arXiv preprint arXiv:2406.19581}, 2024.

\bibitem[De~Luca et~al.(2006)De~Luca, Adam, Wotiz, Gilmore, and
  Nawab]{DeLuca2006}
C.~J. De~Luca, A.~Adam, R.~Wotiz, L.~D. Gilmore, and S.~H. Nawab.
\newblock Decomposition of surface {EMG} signals.
\newblock \emph{Journal of Neurophysiology}, 96\penalty0 (3):\penalty0
  1646--1657, 2006.

\bibitem[Del~Buono et~al.(2019)Del~Buono, Arena, Borlaug, Carbone, Canada,
  Kirkman, Garten, Rodriguez-Miguelez, Guazzi, Lavie, et~al.]{DelBuono2019}
M.~G. Del~Buono, R.~Arena, B.~A. Borlaug, S.~Carbone, J.~M. Canada, D.~L.
  Kirkman, R.~Garten, P.~Rodriguez-Miguelez, M.~Guazzi, C.~J. Lavie, et~al.
\newblock Exercise intolerance in patients with heart failure: Jacc
  state-of-the-art review.
\newblock \emph{Journal of the American College of Cardiology}, 73\penalty0
  (17):\penalty0 2209--2225, 2019.

\bibitem[Del~Vecchio et~al.(2020)Del~Vecchio, Holobar, Falla, Felici, Enoka,
  and Farina]{DelVecchio2020}
A.~Del~Vecchio, A.~Holobar, D.~Falla, F.~Felici, R.~Enoka, and D.~Farina.
\newblock Tutorial: Analysis of motor unit discharge characteristics from
  high-density surface {EMG} signals.
\newblock \emph{Journal of Electromyography and Kinesiology}, 53:\penalty0
  102426, 2020.

\bibitem[Enoka(2019)]{Enoka2019}
R.~M. Enoka.
\newblock Physiological validation of the decomposition of surface {EMG}
  signals.
\newblock \emph{Journal of Electromyography and Kinesiology}, 46:\penalty0
  70--83, 2019.

\bibitem[Farina and Holobar(2016)]{Farina2016a}
D.~Farina and A.~Holobar.
\newblock Characterization of human motor units from surface {EMG}
  decomposition.
\newblock \emph{Proceedings of the IEEE}, 104\penalty0 (2):\penalty0 353--373,
  2016.

\bibitem[Farina and Merletti(2001)]{Farina2001}
D.~Farina and R.~Merletti.
\newblock A novel approach for precise simulation of the {EMG} signal detected
  by surface electrodes.
\newblock \emph{IEEE transactions on biomedical engineering}, 48\penalty0
  (6):\penalty0 637--646, 2001.

\bibitem[Farina et~al.(2008)Farina, Negro, Gazzoni, and Enoka]{Farina2008}
D.~Farina, F.~Negro, M.~Gazzoni, and R.~M. Enoka.
\newblock Detecting the unique representation of motor-unit action potentials
  in the surface electromyogram.
\newblock \emph{Journal of Neurophysiology}, 100\penalty0 (3):\penalty0
  1223--1233, 2008.

\bibitem[Farina et~al.(2014)Farina, Negro, and Dideriksen]{Farina2014}
D.~Farina, F.~Negro, and J.~L. Dideriksen.
\newblock The effective neural drive to muscles is the common synaptic input to
  motor neurons.
\newblock \emph{The Journal of physiology}, 592\penalty0 (16):\penalty0
  3427--3441, 2014.

\bibitem[Formento et~al.(2021)Formento, Botros, and Carmena]{Formento2021}
E.~Formento, P.~Botros, and J.~M. Carmena.
\newblock Skilled independent control of individual motor units via a
  non-invasive neuromuscular--machine interface.
\newblock \emph{Journal of Neural Engineering}, 18\penalty0 (6):\penalty0
  066019, 2021.

\bibitem[Fuglevand et~al.(1993)Fuglevand, Winter, and Patla]{Fuglevand1993}
A.~J. Fuglevand, D.~A. Winter, and A.~E. Patla.
\newblock Models of recruitment and rate coding organization in motor-unit
  pools.
\newblock \emph{Journal of Neurophysiology}, 70\penalty0 (6):\penalty0
  2470--2488, 1993.

\bibitem[Gherardi et~al.(2019)Gherardi, Cr{\'e}peaux, and
  Authier]{Gherardi2019}
R.~K. Gherardi, G.~Cr{\'e}peaux, and F.-J. Authier.
\newblock Myalgia and chronic fatigue syndrome following immunization:
  macrophagic myofasciitis and animal studies support linkage to aluminum
  adjuvant persistency and diffusion in the immune system.
\newblock \emph{Autoimmunity Reviews}, 18\penalty0 (7):\penalty0 691--705,
  2019.

\bibitem[Glaser and Holobar(2018)]{Glaser2018}
V.~Glaser and A.~Holobar.
\newblock Motor unit identification from high-density surface electromyograms
  in repeated dynamic muscle contractions.
\newblock \emph{IEEE Transactions on Neural Systems and Rehabilitation
  Engineering}, 27\penalty0 (1):\penalty0 66--75, 2018.

\bibitem[Goodlich et~al.(2023)Goodlich, Del~Vecchio, and
  Kavanagh]{Goodlich2023}
B.~I. Goodlich, A.~Del~Vecchio, and J.~J. Kavanagh.
\newblock Motor unit tracking using blind source separation filters and
  waveform cross-correlations: reliability under physiological and
  pharmacological conditions.
\newblock \emph{Journal of Applied Physiology}, 135\penalty0 (2):\penalty0
  362--374, 2023.

\bibitem[Grison et~al.(2024)Grison, Clarke, Muceli, Ib{\'a}{\~n}ez, Kundu, and
  Farina]{Grison2024}
A.~Grison, A.~K. Clarke, S.~Muceli, J.~Ib{\'a}{\~n}ez, A.~Kundu, and D.~Farina.
\newblock A particle swarm optimised independence estimator for blind source
  separation of neurophysiological time series.
\newblock \emph{IEEE Transactions on Biomedical Engineering}, 2024.

\bibitem[Grison et~al.(2025)Grison, Mendez~Guerra, Clarke, Muceli,
  Ib{\'a}{\~n}ez, and Farina]{Grison2025}
A.~Grison, I.~Mendez~Guerra, A.~K. Clarke, S.~Muceli, J.~Ib{\'a}{\~n}ez, and
  D.~Farina.
\newblock Unlocking the full potential of high-density surface {EMG}: novel
  non-invasive high-yield motor unit decomposition.
\newblock \emph{The Journal of Physiology}, 603\penalty0 (8):\penalty0
  2281--2300, 2025.

\bibitem[Guerra et~al.(2024)Guerra, Barsakcioglu, and Farina]{Guerra2024}
I.~M. Guerra, D.~Y. Barsakcioglu, and D.~Farina.
\newblock Adaptive {EMG} decomposition in dynamic conditions based on online
  learning metrics with tunable hyperparameters.
\newblock \emph{Journal of Neural Engineering}, 21\penalty0 (4):\penalty0
  046023, 2024.

\bibitem[Heckman et~al.(2008)Heckman, Johnson, Mottram, and
  Schuster]{Heckman2008}
C.~Heckman, M.~Johnson, C.~Mottram, and J.~Schuster.
\newblock Persistent inward currents in spinal motoneurons and their influence
  on human motoneuron firing patterns.
\newblock \emph{The Neuroscientist}, 14\penalty0 (3):\penalty0 264--275, 2008.

\bibitem[Heckman and Enoka(2012)]{Heckman2012}
C.~J. Heckman and R.~M. Enoka.
\newblock Motor {U}nit.
\newblock \emph{Comprehensive Physiology}, 2:\penalty0 2629--2682, 2012.

\bibitem[Hellwig et~al.(2003)Hellwig, Schelter, Guschlbauer, Timmer, and
  L{\"u}cking]{Hellwig2003}
B.~Hellwig, B.~Schelter, B.~Guschlbauer, J.~Timmer, and C.~L{\"u}cking.
\newblock Dynamic synchronisation of central oscillators in essential tremor.
\newblock \emph{Clinical Neurophysiology}, 114\penalty0 (8):\penalty0
  1462--1467, 2003.

\bibitem[Henneman et~al.(1965)Henneman, Somjen, and Carpenter]{Henneman1954}
E.~Henneman, G.~Somjen, and D.~O. Carpenter.
\newblock Excitability and inhibitibility of motoneurons of different sizes.
\newblock \emph{Journal of Neurophysiology}, 28\penalty0 (3):\penalty0
  599--620, 1965.

\bibitem[Holobar and Zazula(2007{\natexlab{a}})]{Holobar2007a}
A.~Holobar and D.~Zazula.
\newblock Multichannel blind source separation using convolution kernel
  compensation.
\newblock \emph{IEEE Transactions on Signal Processing}, 55\penalty0
  (9):\penalty0 4487--4496, 2007{\natexlab{a}}.

\bibitem[Holobar and Zazula(2007{\natexlab{b}})]{Holobar2007b}
A.~Holobar and D.~Zazula.
\newblock Gradient convolution kernel compensation applied to surface
  electromyograms.
\newblock In \emph{International Conference on Independent Component Analysis
  and Signal Separation}, pages 617--624. Springer, 2007{\natexlab{b}}.

\bibitem[Holobar et~al.(2012)Holobar, Glaser, Gallego, Dideriksen, and
  Farina]{Holobar2012}
A.~Holobar, V.~Glaser, J.~A. Gallego, J.~L. Dideriksen, and D.~Farina.
\newblock Non-invasive characterization of motor unit behaviour in pathological
  tremor.
\newblock \emph{Journal of Neural Engineering}, 9\penalty0 (5):\penalty0
  056011, 2012.

\bibitem[Holobar et~al.(2014)Holobar, Minetto, and Farina]{Holobar2014}
A.~Holobar, M.~A. Minetto, and D.~Farina.
\newblock Accurate identification of motor unit discharge patterns from
  high-density surface {EMG} and validation with a novel signal-based
  performance metric.
\newblock \emph{Journal of Neural Engineering}, 11\penalty0 (1):\penalty0
  016008, 2014.

\bibitem[Hug et~al.(2025)Hug, Dernoncourt, Avrillon, Thorstensen, Besomi,
  van~den Hoorn, and Tucker]{Hug2025}
F.~Hug, F.~Dernoncourt, S.~Avrillon, J.~Thorstensen, M.~Besomi, W.~van~den
  Hoorn, and K.~Tucker.
\newblock Non-homogeneous distribution of inhibitory inputs among motor units
  in response to nociceptive stimulation at moderate contraction intensity.
\newblock \emph{The Journal of Physiology}, 2025.

\bibitem[Hussey and Gupta(2022)]{Hussey2022}
C.~Hussey and A.~Gupta.
\newblock Exercise interventions to combat cancer-related fatigue in cancer
  patients undergoing treatment: a review.
\newblock \emph{Cancer Investigation}, 40\penalty0 (9):\penalty0 822--838,
  2022.

\bibitem[Hyv{\"a}rinen and Oja(2000)]{Hyvarinen2000}
A.~Hyv{\"a}rinen and E.~Oja.
\newblock Independent component analysis: algorithms and applications.
\newblock \emph{Neural networks}, 13\penalty0 (4-5):\penalty0 411--430, 2000.

\bibitem[Hyv\"{a}rinen et~al.(1998{\natexlab{a}})Hyv\"{a}rinen, Hoyer, and
  Oja]{Hyvarinen1998a}
A.~Hyv\"{a}rinen, P.~Hoyer, and E.~Oja.
\newblock Sparse code shrinkage for image denoising.
\newblock In \emph{1998 IEEE International Joint Conference on Neural Networks
  Proceedings. IEEE World Congress on Computational Intelligence (Cat.
  No.98CH36227)}, volume~2, pages 859--864, 1998{\natexlab{a}}.
\newblock \doi{10.1109/IJCNN.1998.685880}.

\bibitem[Hyv\"{a}rinen et~al.(1998{\natexlab{b}})Hyv\"{a}rinen, Hoyer, and
  Oja]{Hyvarinen1998b}
A.~Hyv\"{a}rinen, P.~Hoyer, and E.~Oja.
\newblock Sparse code shrinkage: Denoising by nonlinear maximum likelihood
  estimation.
\newblock In M.~Kearns, S.~Solla, and D.~Cohn, editors, \emph{Advances in
  Neural Information Processing Systems}, volume~11. MIT Press,
  1998{\natexlab{b}}.

\bibitem[Hyv\"arinen et~al.(2001)Hyv\"arinen, Karhunen, and Oja]{Hyvarinen2001}
A.~Hyv\"arinen, J.~Karhunen, and E.~Oja.
\newblock \emph{Independent {Component} {Analysis}}.
\newblock John Wiley \& Sons, Ltd, 2001.
\newblock ISBN 978-0-471-22131-9.

\bibitem[Klotz et~al.(2020)Klotz, Gizzi, Yavuz, and R{\"o}hrle]{Klotz2020}
T.~Klotz, L.~Gizzi, U.~{\c{S}}. Yavuz, and O.~R{\"o}hrle.
\newblock Modelling the electrical activity of skeletal muscle tissue using a
  multi-domain approach.
\newblock \emph{Biomechanics and modeling in mechanobiology}, 19\penalty0
  (1):\penalty0 335--349, 2020.

\bibitem[Klotz et~al.(2023)Klotz, Lehmann, Negro, and R{\"o}hrle]{Klotz2023}
T.~Klotz, L.~Lehmann, F.~Negro, and O.~R{\"o}hrle.
\newblock High-density magnetomyography is superior to high-density surface
  electromyography for motor unit decomposition: a simulation study.
\newblock \emph{Journal of Neural Engineering}, 20\penalty0 (4):\penalty0
  046022, 2023.

\bibitem[Kramberger and Holobar(2021)]{Kramberger2021}
M.~Kramberger and A.~Holobar.
\newblock On the prediction of motor unit filter changes in blind source
  separation of high-density surface electromyograms during dynamic muscle
  contractions.
\newblock \emph{IEEE Access}, 9:\penalty0 103533--103540, 2021.

\bibitem[Krizhevsky(2009)]{Krizhevsky2009}
A.~Krizhevsky.
\newblock Learning multiple layers of features from tiny images.
\newblock Technical report, Toronto, ON, Canada, 2009.

\bibitem[Lin et~al.(2024)Lin, Cui, Chen, Liu, and Jiang]{Lin2024}
C.~Lin, Z.~Cui, C.~Chen, Y.~Liu, and N.~Jiang.
\newblock A fast gradient convolution kernel compensation method for surface
  electromyogram decomposition.
\newblock \emph{Journal of Electromyography and Kinesiology}, 76:\penalty0
  102869, 2024.

\bibitem[Lowery et~al.(2002)Lowery, Stoykov, Taflove, and Kuiken]{Lowery2002}
M.~M. Lowery, N.~S. Stoykov, A.~Taflove, and T.~A. Kuiken.
\newblock A multiple-layer finite-element model of the surface {EMG} signal.
\newblock \emph{IEEE Transactions on Biomedical Engineering}, 49\penalty0
  (5):\penalty0 446--454, 2002.

\bibitem[Lubel et~al.(2024)Lubel, Rohl\'en, Sgambato, Barsakcioglu, Ibanez,
  Tang, and Farina]{Lubel2024}
E.~Lubel, R.~Rohl\'en, B.~G. Sgambato, D.~Y. Barsakcioglu, J.~Ibanez, M.-X.
  Tang, and D.~Farina.
\newblock Accurate identification of motoneuron discharges from ultrasound
  images across the full muscle cross-section.
\newblock \emph{IEEE Transactions on Biomedical Engineering}, 71\penalty0
  (5):\penalty0 1466--1477, 2024.

\bibitem[Lundsberg et~al.(2023)Lundsberg, Bj{\"o}rkman, Malesevic, and
  Antfolk]{Lundsberg2023}
J.~Lundsberg, A.~Bj{\"o}rkman, N.~Malesevic, and C.~Antfolk.
\newblock Compressed spike-triggered averaging in iterative decomposition of
  surface {EMG}.
\newblock \emph{Computer Methods and Programs in Biomedicine}, 228:\penalty0
  107250, 2023.

\bibitem[McGill et~al.(2005)McGill, Lateva, and Marateb]{McGill2005}
K.~C. McGill, Z.~C. Lateva, and H.~R. Marateb.
\newblock {EMGLAB}: an interactive {EMG} decomposition program.
\newblock \emph{Journal of neuroscience methods}, 149\penalty0 (2):\penalty0
  121--133, 2005.

\bibitem[McNulty et~al.(2000)McNulty, Falland, and Macefield]{McNulty2000}
P.~McNulty, K.~Falland, and V.~Macefield.
\newblock Comparison of contractile properties of single motor units in human
  intrinsic and extrinsic finger muscles.
\newblock \emph{The Journal of Physiology}, 526\penalty0 (2):\penalty0
  445--456, 2000.

\bibitem[Meng et~al.(2022)Meng, Chen, Jiang, Liu, Fan, Dai, and Chen]{Meng2022}
L.~Meng, Q.~Chen, X.~Jiang, X.~Liu, J.~Fan, C.~Dai, and W.~Chen.
\newblock Evaluation of decomposition parameters for high-density surface
  electromyogram using fast independent component analysis algorithm.
\newblock \emph{Biomedical Signal Processing and Control}, 75:\penalty0 103615,
  2022.

\bibitem[Merletti et~al.(2008)Merletti, Holobar, and Farina]{Merletti2008}
R.~Merletti, A.~Holobar, and D.~Farina.
\newblock Analysis of motor units with high-density surface electromyography.
\newblock \emph{Journal of Electromyography and Kinesiology}, 18\penalty0
  (6):\penalty0 879--890, 2008.

\bibitem[Mesquita et~al.(2024)Mesquita, Taylor, Heckman, Trajano, and
  Blazevich]{Mesquita2024}
R.~N. Mesquita, J.~L. Taylor, C.~Heckman, G.~S. Trajano, and A.~J. Blazevich.
\newblock Persistent inward currents in human motoneurons: Emerging evidence
  and future directions.
\newblock \emph{Journal of Neurophysiology}, 132\penalty0 (4):\penalty0
  1278--1301, 2024.

\bibitem[Montes et~al.(2021)Montes, Goodwin, McDermott, Uher, Hernandez,
  Coutts, Cocchi, Hauschildt, Cornett, Rao, et~al.]{Montes2021}
J.~Montes, A.~M. Goodwin, M.~P. McDermott, D.~Uher, F.~M. Hernandez, K.~Coutts,
  J.~Cocchi, M.~Hauschildt, K.~M. Cornett, A.~K. Rao, et~al.
\newblock Diminished muscle oxygen uptake and fatigue in spinal muscular
  atrophy.
\newblock \emph{Annals of Clinical and Translational Neurology}, 8\penalty0
  (5):\penalty0 1086--1095, 2021.

\bibitem[Mr{\'o}wczy{\'n}ski et~al.(2015)Mr{\'o}wczy{\'n}ski, Celichowski,
  Raikova, and Krutki]{Mrowczynski2015}
W.~Mr{\'o}wczy{\'n}ski, J.~Celichowski, R.~Raikova, and P.~Krutki.
\newblock Physiological consequences of doublet discharges on motoneuronal
  firing and motor unit force.
\newblock \emph{Frontiers in Cellular Neuroscience}, 9:\penalty0 81, 2015.

\bibitem[Mukamel et~al.(2009)Mukamel, Nimmerjahn, and Schnitzer]{Mukamel2009}
E.~A. Mukamel, A.~Nimmerjahn, and M.~J. Schnitzer.
\newblock Automated analysis of cellular signals from large-scale calcium
  imaging data.
\newblock \emph{Neuron}, 63\penalty0 (6):\penalty0 747--760, 2009.

\bibitem[Nawab et~al.(2010)Nawab, Chang, and De~Luca]{Nawab2010}
S.~H. Nawab, S.-S. Chang, and C.~J. De~Luca.
\newblock High-yield decomposition of surface {EMG} signals.
\newblock \emph{Clinical Neurophysiology}, 121\penalty0 (10):\penalty0
  1602--1615, 2010.

\bibitem[Negro and Farina(2011)]{Negro2011}
F.~Negro and D.~Farina.
\newblock Decorrelation of cortical inputs and motoneuron output.
\newblock \emph{Journal of Neurophysiology}, 106\penalty0 (5):\penalty0
  2688--2697, 2011.

\bibitem[Negro et~al.(2016)Negro, Muceli, Castronovo, Holobar, and
  Farina]{Negro2016}
F.~Negro, S.~Muceli, A.~M. Castronovo, A.~Holobar, and D.~Farina.
\newblock Multi-channel intramuscular and surface {EMG} decomposition by
  convolutive blind source separation.
\newblock \emph{Journal of Neural Engineering}, 13\penalty0 (2):\penalty0
  026027, 2016.

\bibitem[Oliveira and Negro(2021)]{Oliveira2021}
A.~S. Oliveira and F.~Negro.
\newblock Neural control of matched motor units during muscle shortening and
  lengthening at increasing velocities.
\newblock \emph{Journal of Applied Physiology}, 130\penalty0 (6):\penalty0
  1798--1813, 2021.

\bibitem[Piotrkiewicz et~al.(2013)Piotrkiewicz, Sebik, Binbo{\u{g}}a,
  M{\l}o{\'z}niak, Kuraszkiewicz, and T{\"u}rker]{Piotrkiewicz2013}
M.~Piotrkiewicz, O.~Sebik, E.~Binbo{\u{g}}a, D.~M{\l}o{\'z}niak,
  B.~Kuraszkiewicz, and K.~S. T{\"u}rker.
\newblock Double discharges in human soleus muscle.
\newblock \emph{Frontiers in Human Neuroscience}, 7:\penalty0 843, 2013.

\bibitem[Powers et~al.(2012)Powers, ElBasiouny, Rymer, and Heckman]{Powers2012}
R.~K. Powers, S.~M. ElBasiouny, W.~Z. Rymer, and C.~J. Heckman.
\newblock Contribution of intrinsic properties and synaptic inputs to
  motoneuron discharge patterns: a simulation study.
\newblock \emph{Journal of neurophysiology}, 107\penalty0 (3):\penalty0
  808--823, 2012.

\bibitem[Raikova et~al.(2007)Raikova, Celichowski, Pogrzebna, Aladjov, and
  Krutki]{Raikova2007}
R.~Raikova, J.~Celichowski, M.~Pogrzebna, H.~Aladjov, and P.~Krutki.
\newblock Modeling of summation of individual twitches into unfused tetanus for
  various types of rat motor units.
\newblock \emph{Journal of Electromyography and Kinesiology}, 17\penalty0
  (2):\penalty0 121--130, 2007.

\bibitem[Raikova et~al.(2008)Raikova, Pogrzebna, Drzyma{\l}a, Celichowski, and
  Aladjov]{Raikova2008}
R.~Raikova, M.~Pogrzebna, H.~Drzyma{\l}a, J.~Celichowski, and H.~Aladjov.
\newblock Variability of successive contractions subtracted from unfused
  tetanus of fast and slow motor units.
\newblock \emph{Journal of Electromyography and Kinesiology}, 18\penalty0
  (5):\penalty0 741--751, 2008.

\bibitem[Rohl{\'e}n and Klotz(2025)]{KlotzRohlen2025}
R.~Rohl{\'e}n and T.~Klotz.
\newblock {Replication data for "Revisiting convolutive blind source separation
  for identifying spiking motor neuron activity: From theory to practice"},
  2025.
\newblock URL \url{https://doi.org/10.5281/zenodo.14824963}.

\bibitem[Rohl{\'e}n et~al.(2020)Rohl{\'e}n, St{\aa}lberg, and
  Gr{\"o}nlund]{Rohlen2020}
R.~Rohl{\'e}n, E.~St{\aa}lberg, and C.~Gr{\"o}nlund.
\newblock Identification of single motor units in skeletal muscle under low
  force isometric voluntary contractions using ultrafast ultrasound.
\newblock \emph{Scientific Reports}, 10\penalty0 (1):\penalty0 22382, 2020.

\bibitem[Rohl{\'e}n et~al.(2025)Rohl{\'e}n, Lubel, and Farina]{Rohlen2025}
R.~Rohl{\'e}n, E.~Lubel, and D.~Farina.
\newblock Assessing the impact of degree of fusion and muscle fibre twitch
  shape variation on the accuracy of motor unit discharge time identification
  from ultrasound images.
\newblock \emph{Biomedical Signal Processing and Control}, 100:\penalty0
  107002, 2025.

\bibitem[R{\"o}hrle et~al.(2019)R{\"o}hrle, Yavuz, Klotz, Negro, and
  Heidlauf]{Rohrle2019}
O.~R{\"o}hrle, U.~{\c{S}}. Yavuz, T.~Klotz, F.~Negro, and T.~Heidlauf.
\newblock Multiscale modeling of the neuromuscular system: coupling
  neurophysiology and skeletal muscle mechanics.
\newblock \emph{Wiley Interdisciplinary Reviews: Systems Biology and Medicine},
  11\penalty0 (6):\penalty0 e1457, 2019.
\newblock \doi{10.1002/wsbm.1457}.

\bibitem[Schmid et~al.(2024)Schmid, Klotz, R{\"o}hrle, Powers, Negro, and
  Yavuz]{Schmid2024}
L.~Schmid, T.~Klotz, O.~R{\"o}hrle, R.~K. Powers, F.~Negro, and U.~{\c{S}}.
  Yavuz.
\newblock Postinhibitory excitation in motoneurons can be facilitated by
  hyperpolarization-activated inward currents: A simulation study.
\newblock \emph{PLOS Computational Biology}, 20\penalty0 (1):\penalty0
  e1011487, 2024.

\bibitem[Silva et~al.(2022)Silva, Bichara, Carneiro, Palacios, Berg, Quaresma,
  and Magno~Falc{\~a}o]{Silva2022}
C.~C. Silva, C.~N.~C. Bichara, F.~R.~O. Carneiro, V.~R. d. C.~M. Palacios,
  A.~V. S. V.~d. Berg, J.~A.~S. Quaresma, and L.~F. Magno~Falc{\~a}o.
\newblock Muscle dysfunction in the long coronavirus disease 2019 syndrome:
  Pathogenesis and clinical approach.
\newblock \emph{Reviews in Medical Virology}, page e2355, 2022.

\bibitem[St{\aa}lberg(2011)]{Stalberg2011}
E.~St{\aa}lberg.
\newblock Macro electromyography, an update.
\newblock \emph{Muscle \& Nerve}, 44\penalty0 (2):\penalty0 292--302, 2011.

\bibitem[Valli et~al.(2024)Valli, Ritsche, Casolo, Negro, and
  De~Vito]{Valli2024}
G.~Valli, P.~Ritsche, A.~Casolo, F.~Negro, and G.~De~Vito.
\newblock Tutorial: Analysis of central and peripheral motor unit properties
  from decomposed high-density surface {EMG} signals with openhdemg.
\newblock \emph{Journal of Electromyography and Kinesiology}, 74:\penalty0
  102850, 2024.

\bibitem[Wu et~al.(2021)Wu, Ditroilo, Delahunt, and De~Vito]{Wu2021}
R.~Wu, M.~Ditroilo, E.~Delahunt, and G.~De~Vito.
\newblock Age related changes in motor function (ii). decline in motor
  performance outcomes.
\newblock \emph{International Journal of Sports Medicine}, 42\penalty0
  (03):\penalty0 215--226, 2021.

\bibitem[Xia et~al.(2024)Xia, Chen, Sheng, and Ding]{Xia2024}
M.~Xia, C.~Chen, X.~Sheng, and H.~Ding.
\newblock Integration of motor unit filters for enhanced surface electromyogram
  decomposition during varying force isometric contraction.
\newblock \emph{IEEE Transactions on Neural Systems and Rehabilitation
  Engineering}, 2024.

\bibitem[Yokoyama et~al.(2021)Yokoyama, Sasaki, Kaneko, Saito, and
  Nakazawa]{Yokoyama2021}
H.~Yokoyama, A.~Sasaki, N.~Kaneko, A.~Saito, and K.~Nakazawa.
\newblock Robust identification of motor unit discharges from high-density
  surface {EMG} in dynamic muscle contractions of the tibialis anterior.
\newblock \emph{IEEE Access}, 9:\penalty0 123901--123911, 2021.

\bibitem[Zheng et~al.(2023)Zheng, Ma, Liu, Houston, Guo, Lian, Li, Zhou, and
  Zhang]{Zheng2023}
Y.~Zheng, Y.~Ma, Y.~Liu, M.~Houston, C.~Guo, Q.~Lian, S.~Li, P.~Zhou, and
  Y.~Zhang.
\newblock High-density surface {EMG} decomposition by combining iterative
  convolution kernel compensation with an energy-specific peel-off strategy.
\newblock \emph{IEEE Transactions on Neural Systems and Rehabilitation
  Engineering}, 2023.

\bibitem[Zhou and Rymer(2004)]{Zhou2004}
P.~Zhou and W.~Z. Rymer.
\newblock {MUAP} number estimates in surface {EMG}: template-matching methods
  and their performance boundaries.
\newblock \emph{Annals of Biomedical Engineering}, 32:\penalty0 1007--1015,
  2004.

\end{thebibliography}

\end{document}